\documentclass[]{aa}
\usepackage{afterpage}
\usepackage{amsmath}
\usepackage{dcolumn}
\usepackage{graphicx}
\usepackage{longtable}
\usepackage{lscape}
\usepackage{graphicx}
\usepackage{natbib}
\usepackage{afterpage,amssymb,booktabs,dcolumn,longtable,lscape,palatino,rotating,subfigure,supertabular,txfonts}
\newcommand{\htwoo}{H$_2$O}
\newcommand{\methanol}{CH$_3$OH}

\newcommand{\s}{s$^{-1}$}
\begin{document}

\title{OH Masers and the Dust Emissions Towards HMPOs}

\author{K. A. Edris}

\institute{
Astronomy Dept., Faculty of science, Al-Azhar University, Naser City, Cairo, Egypt}

\authorrunning{K. A. Edris}
\offprints{khedres@azhar.edu.eg}
\large
\abstract{

\textit{Context}. OH maser emission is known to be associated with high mass star forming regions. Towards some of these regions, OH masers are associated with HII regions. Towards others, believed to be in an earlier evolutionary state, OH masers are offset from HII regions. Towards these later regions, it is believed that OH masers are associated with the circumstellar disk (e.g. Edris et al. \citealp{edris05}; Gray et al. \citealp{gray03}). These disks should be hosting dense dust grains. The presence of the hot dust could be traced via the millimeter continuum emission as well as IR emission.

\textit{Aims}. studying the association between millimeter (mm) continuum, the OH masers emission, and IRAS sources.

\textit{Methods}. A sample of 27 High Mass Star Forming Regions (HMSFRs) chosen from IRAS catalog and show OH maser emission (Edris et. al. \citealp{edris07}) have been studied at 1.1 millimeter (mm) continuum emission of the Bolocam Galactic Plane Survey (BGPS).

\textit{Results}. The 1.1-mm continuum emission have been found within $\simeq$ 30$'$ towards 23 sources of the OH maser sample. These sources were divided into three groups depends on the offset of the closest mm peak from the OH maser position. The association between the OH, mm and IR emissions types have been confirmed for two sources. Generally the IRAS position is more consistent with the mm peaks than the OH maser emission and towards 10 sources the IRAS and OH masers are not consistent with the same mm peak.

\textit{Conclusion}. The relatively large positional uncertainty do not allow to firm conclusions but it seems that the IR peak is closer to the mm emission than the OH maser.

  \keywords{stars: formation, masers: OH masers, millimeter} }

\maketitle

\section{Introduction}

Compact HII regions, molecular outflows and circumstellar disks are signs of the existence of massive protostars (e.g. Garay \& Lizano \citealp{garay99}; Churchwell \citealp{churchwell02}). Maser emission is also found to be associated with these objects (e.g. Garay \& Lizano \citealp{garay99}; Edris et al. \citealp{edris05}) with OH one of most widespread types of maser associated with these regions. OH masers are associated with two different stages of star formation. The first one is associated with circumstellar disks and molecular outflows (Cohen, Rowland \& Blair \citealp{cohen84}; Brebner \citealp{brebner88}; Cohen et al. \citealp{cohen03}; Edris et al. \citealp{edris07}). the other one is a relatively advanced star forming stage of the appearance of UCHII region (e.g. Garay \& Lizano \citealp{garay99}, and references therein). The mm emission could trace different phenomena of star formation process. It could trace slowly collapsing rotating molecular fragment onto itself, a circumstellar disk and molecular outflows (e.g. Zinnecker et al. \citealp{zinnecker92} ; Adams et al. 1990 ; Beuther et al \citealp{beuther02b}). therefore the mm emission should be associated with the first type of OH masers mentioned above.
The association between the maser emission and mm and sub-mm have studied in case of methanol masers by Breen et al. (2010) and water masers by Jenness et al. (1995). The association of OH masers and the mm emission has not been studied. With this in mind a sample of high mass protostellar objects has been studied at the mm emission.

\section{The sample}

The sample of sources in this present paper is partially drawn from the surveys of Sridharan et al. (\citealp{sridharan02}) and Molinari et al. (\citealp{molinari96}). These two samples are originally drawn from the IRAS Point Source Catalog based on their colour selection criteria. They are believed to contain massive sources in a very early stage of evolution prior to the forming of UCHII regions. Molinari et al. (\citealp{molinari96}) divided their sample into two types: \textit{High} and \textit{Low} sources. The Sridharan et al. (\citealp{sridharan02}) sample and the \textit{High} sources of Molinari et al (1996) have similar colour. They both satisfy the criteria of Wood \& Churchwell (\citealp{wood89}) for a UCHII region colour. However the sources in these two samples (as well as the Molinari et al. (\citealp{molinari96}) \textit{Low} sources) are not known to be associated with detectable HII regions. Molinari et al. (\citealp{molinari96}) suggest that their \textit{Low} sources comprises objects which are in a different evolutionary stage from those in their High sources, and therefore also from the other sample. Further details about these sources and their selection criteria can be found in Sridharan et al. (2002) and Molinari et al. (1996) and references therein as well as their follow up papers.

Edris, Fuller \& Cohen (\citealp{edris07}, hereafter EFC07) surveyed these two samples for OH maser emission and detected the masers towards 26 \% of the sources. One of these sources (IRAS $20126+4104$) have been imaged in high angular resolution by Edris et al. (\citealp{edris05}). The OH masers are originate from the circumstellar disk rotating around the protostar. This source is closely associated with 1.3 and 3 mm emission (Cesaroni et al. \citealp{cesaroni99}; Cesaroni et al. \citealp{cesaroni97}). The 27 sources studied in this paper is the OH maser sources that have been found to have images at 1.1-mm emission in the BGPS survey. These sample contains 13 sources from Sridharan et al. (\citealp{sridharan02}) sample and 17 sources from Molinari et al. (1996) with three sources common between the two sample. These 17 sources are 11 \textit{High} type sources and 6 \textit{Low} type sources.

\section{Observations and archive data}

\subsection{OH Masers Observations}

The Nan\c{c}ay radio telescope and the 100$-$m single dish Green Bank Telescope (GBT) have been used to observe the four OH transitions at 1665, 1667, 1612 and 1720 MHz in both left and right circular polarizations. After confirming the presence of an OH maser towards the IRAS position of a source, the GBT was used to map the sources small 3 arcmin sampled maps, typically $3\times3$ pixels in size, to determine the position of the peak emission. Further details about these observations can be found in Edris, Fuller, \& Cohen (\citealp{edris07}).

\subsection{Archive Data}

The millimeter data is based on the Bolocam Galactic Plane Survey (BGPS). The used tables have been downloaded from the http://irsa.ipac.caltech.edu/data/. The survey uses the 144-element Bolocam array on the Caltech Submillimeter Observatory, which observes in a band centered at 268 GHz (1.1-mm) and a width of 46 GHz. The bandpass is designed to reject emission from the CO(2,1) transition, which is the dominant line contributor at these wavelengths. Detailed description of the BGPS survey observations and methods can be found at http://irsa.ipac.caltech.edu/data.
%(at %http://irsa.ipac.caltech.edu/data/$BOLOCAM_GPS$/bgps_methods.pdf) and references therein.

\begin{figure}
  \centering
  \includegraphics[angle=-90,width=9cm]{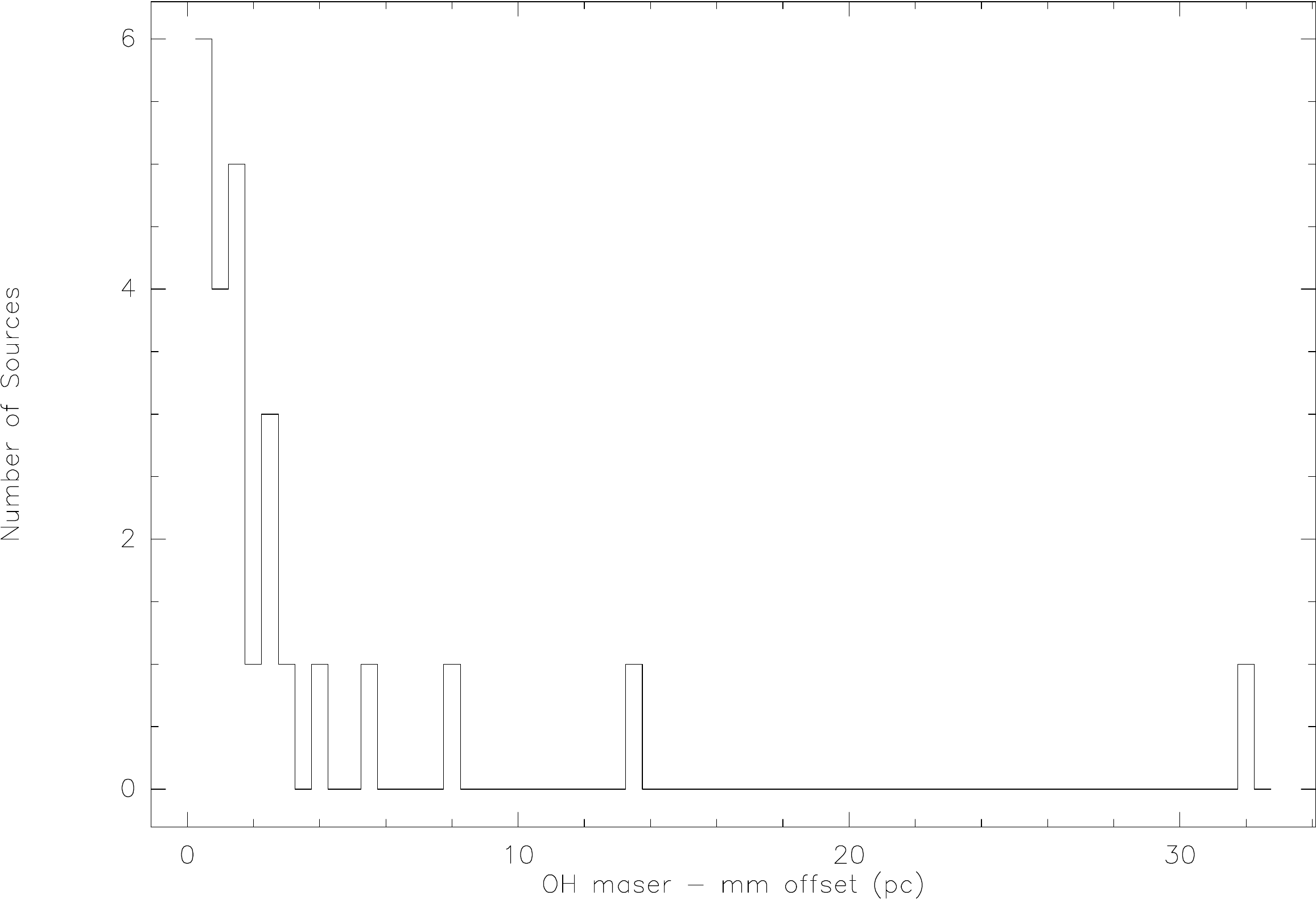}
\caption{The offsets of the OH masers and the closest mm peak.}
\label{fig:offset}
\end{figure}

\section{Results}

Among the 27 OH maser sources, 23 sources have been found to be associated with many peaks of millimeter emission within 30$'$. The offsets of the closest mm peak have been calculated. The offset of $<$ 0.5 pc have been found at 7 sources. The offset of two sources of them are $\leq$ 0.05 pc. These two sources are IRAS $18089$-$1732$ and IRAS $19035+0641$. They represent two different samples of High Mass Protostellar Objects (HMPOs) candidates. IRAS $18089$-$1732$ is one of the \textit{High} type sample of Molinari et al. \citealp{molinari96} while IRAS $19035+0641$ is from the sample of Sridharan et al. \citealp{sridharan02}. The millimeter peak towards $18089$-$1732$ and IRAS $19035+0641$ are probably coincident with the OH maser emission as well as the IRAS position.
The offsets of the next 12 sources are 0.6 $-$ 2 pc. The rest of sources have offsets $\geq$ 3 pc (figure \ref{fig:offset}). The four sources without any close mm emission contain three from the \textit{High} sources of Molinari et al. (1996) and one from the sample of Sridharan et al. (2002).

positions maps for the 23 sources are shown in figure (\ref{fig:1mm} and \ref{fig:2mm}) . Table 1 gives the parameters of the studied sample, namely the source name, the type, the position of the OH maser emission, the position of the closest 1.1-mm peak, the distances of the sources, the offset of the two position in pc and the peak velocities of the OH sources. The OH masers positions are from the GBT observations (Edris, Fuller, \& Cohen 2007); these positions have been drawn from the 1665-MHz line emission. Some sources which show different position for the 1667-MHz line, the position of the 1667-MHz line emission is mentioned as well. Their positions and rms errors have been calculated from 9 points maps. The distances of the sources have been taken from Molinari et al. (\citealp{molinari96}) and Sridharan et al. (\citealp{sridharan02}) except IRAS $19118+0945$ no distance was given by Molinari et al. (\citealp{molinari96}).

\vspace{1cm}

%%%%%%%%%%%%%%%%%%%%%%%%%%%%%%%%%%%%%%%%%%%%%%%%%%%%%%%%%%%%%%%%%%%%%%%%%%%%%%
\onecolumn
%\begin{landscape}
{\scriptsize%
  \begin{center}
    \tablefirsthead{
      \toprule%
      \toprule%
      IRAS Name                              &
      type                                      &
      \multicolumn{2}{c}{OH masers position}       &
      \multicolumn{2}{c}{1.1 mm position}         &
      Distance         &
      Offset         &
      peak velocity                             \\
      \cmidrule(r){3-4}\cmidrule(lr){5-6}
      &
      &
      \multicolumn{1}{c}{RA(J2000)}          &
      \multicolumn{1}{c}{DEC(J2000)}         &

      \multicolumn{1}{c}{RA(J2000)}          &
      \multicolumn{1}{c}{DEC(J2000)}         &
      Kpc &
      arcsec&
      km \/ s      \\
%      \cmidrule(r){4-6}
      &
      &
      \multicolumn{1}{c}{h~~m~~s}                           &
      \multicolumn{1}{c}{$^{\circ}$~~$'$~~$''$}         &

      \multicolumn{1}{c}{h~~m~~s}                           &
      \multicolumn{1}{c}{$^{\circ}$~~$'$~~$''$}         &
      &
      \\
      \midrule
    }
    \tablehead{%
      \toprule%
      \multicolumn{10}{l}{\small\sl continued from previous page}\\
      \toprule%
      IRAS Name                                 &
      Type                                      &
      \multicolumn{2}{c}{OH masers position}       &
      \multicolumn{3}{c}{1.1 mm position}         &
      \multicolumn{1}{c}{Distance}         &
      \multicolumn{1}{c}{Offset}         &

      R                                      &
      R                                      &
      R                                      &
      Ref./Notes                            \\
      \cmidrule(r){2-3}\cmidrule(lr){4-5}
      &
      \multicolumn{1}{c}{RA(J2000)}       &
      \multicolumn{1}{c}{DEC(J2000)}         &
      &
      \multicolumn{1}{c}{RA(J2000)}       &
      \multicolumn{1}{c}{DEC(J2000)}         &
      \multicolumn{1}{c}{}         &
      &
      &
      \\
      \cmidrule(r){5-7}
      &
      \multicolumn{1}{c}{kpc}                           &
      \multicolumn{1}{c}{$^{\circ}$ ~ $'$ ~ $''$}         &
      &
      \multicolumn{1}{c}{h~~m~~s}                           &
      \multicolumn{1}{c}{$^{\circ}$ ~ $'$ ~ $''$}         &
      \multicolumn{1}{c}{$''$}         &
      &
      \\
      \midrule
    }
    \tabletail{%
      \midrule
      \multicolumn{10}{r}{\small\sl continued on next page}\\
      \midrule}
    \tablelasttail{\bottomrule}
  \tablecaption{The OH masers positions and the positions of the closest 1.1 mm emission. The OH positions are measured from GBT observations (Edris, Fuller, \& Cohen 2007) except for sources which have higher resolution observations their position have been taken from. These sources are IRAS $18089$-$1732$ and IRAS $19035+0641$ (Argon et al. 2000) and IRAS $19092+0841$ (Edris et al. 2009).}
\begin{supertabular}{lccccccccc}

\object{06056+2131}   & H   & 06 08 52.4 $\pm$ 1.6    &     21 34 06  $\pm$  23  & 06 08 53  &      21 35 16  & 1.5  &   70  &   10.14   &        \\
\object{17527$-$2439} & H   & 17 55 28.3 $\pm$ 1.9    &  $-$24 36 36  $\pm$  27  & 17 55 33  &   $-$24 41 52  & 3.23 &  323  &   11.53   &      \\
\object{18024$-$2119} & L   & 18 05 25.6 $\pm$ 2.3    &  $-$21 14 59  $\pm$  19  & 18 05 26  &   $-$21 19 24  & 0.12 &  264  &  $-$4.05  &      \\
\object{18048$-$2019} & H   & 18 07 44.6 $\pm$ 2.2    &  $-$20 18 41  $\pm$  38  & 18 07 46  &   $-$20 19 47  & 4.99 &   68  &   44.36   &       \\
\object{18089$-$1732} & H/S & 18 11 51.4 $\pm$ 0.0    &  $-$17 31 29  $\pm$  01  & 18 11 51  &   $-$17 31 26  & 3.6  &    3  &   32.92   &        \\
\object{18090$-$1832} & S   & 18 11 47.4 $\pm$ 1.5    &  $-$18 29 47  $\pm$  26  & 18 11 44  &   $-$18 43 14  & 6.6  &  808  &  108.9    &      \\
\object{18102$-$1800} & S   & 18 13 04.4 $\pm$ 1.5    &  $-$18 00 23  $\pm$  16  & 18 13 11  &   $-$17 59 48  & 2.6  &  110  &   24.40   &      \\
\object{18144$-$1723} & L   & 18 17 26.5 $\pm$ 1.1    &  $-$17 22 29  $\pm$  16  & 18 17 15  &   $-$17 12 08  & 4.33 &  644  &   48.33   &      \\
\object{18182$-$1433} & S   & 18 21 11.0 $\pm$ 1.0    &  $-$14 31 23  $\pm$  14  & 18 21 09  &   $-$14 31 45  & 4.5  &   22  &   61.55   &       \\
\object{18236$-$1205} & H   & 18 26 36.2 $\pm$ 1.0    &  $-$12 04 54  $\pm$  14  & 18 26 39  &   $-$12 04 53  & 2.51 &   48  &   31.09   &       \\
                      &     & 18 26 31.7 $\pm$ 1.0    &  $-$12 03 26  $\pm$  14  & 18 26 26  &   $-$12 03 57  & 2.51 & 92  &   62.45   &       \\
\object{18278$-$1009} & L   & 18 30 37.9 $\pm$ 1.1    &  $-$10 07 25  $\pm$  17  & 18 30 37  &   $-$10 07 57  & 5.7  &   33  &  119.7    &       \\
\object{18290$-$0924} & S   & 18 31 46.4 $\pm$ 1.1    &  $-$09 22 14  $\pm$  15  & 18 31 43  &   $-$09 22 20  & 5.3  &   43  &   78.33   &       \\
\object{18360$-$0537} & H   & 18 38 42.2 $\pm$ 1.1    &  $-$05 36 25  $\pm$  16  & 18 38 48  &   $-$05 36 20  & 6.28 &   95  &  102.9    &       \\
                      &     & 18 38 47.2 $\pm$ 1.2    &  $-$05 35 40  $\pm$  17  & 18 38 48  &   $-$05 36 20  & 6.28 &   44  &  105.3    &     \\
\object{18385$-$0512} & S   & 18 41 18.2 $\pm$ 1.0    &  $-$05 08 57  $\pm$  14  & 18 41 19  &   $-$05 08 11  & 2    &   49  &   24.87   &       \\
\object{18440$-$0148} & S   & 18 46 37.8 $\pm$ 1.0    &  $-$01 44 27  $\pm$  13  & 18 46 36  &   $-$01 45 15  & 8.3  &   54  &  101.4    &       \\
\object{18454$-$0158} & S   & 18 48 01.3 $\pm$ 1.0    &  $-$01 54 34  $\pm$  15  & 18 48 01  &   $-$01 53 47  & 5.6  &   47  &   39.6    &       \\
\object{18488+0000}   & H/S & 18 51 30.5 $\pm$ 1.0    &     00 03 21  $\pm$  16  & 18 51 36  &      00 02 58  & 5.4  &   92  &   79.57   &       \\
\object{18527+0301}   & L   & 18 54 46.5 $\pm$ 2.5    &     03 05 07  $\pm$  40  & 18 56 03  &      02 56 54  & 5.26 & 1252  &   74.44   &     \\
\object{18566+0408}   & H/S & 18 59 08.6 $\pm$ 1.1    &     04 10 21  $\pm$  17  & 18 59 10  &      04 12 18  & 6.7  &  119  &   83.41   &      \\
                      &     & 18 59 10.4 $\pm$ 1.0    &     04 13 21  $\pm$  14  & 18 59 10  &      04 12 18  & 6.7  &   63  &   81.52   &       \\
\object{19035+0641}   & S   & 19 06 01.6 $\pm$ 0.0    &     06 46 35  $\pm$  01  & 19 06 02  &      06 46 37  & 2.2  &    4  &   32.44   &        \\
\object{19092+0841}   & L   & 19 11 45.9 $\pm$ 0.4    &     08 46 49  $\pm$  06  & 19 11 39  &      08 46 30  & 4.48 &    8  &   57.87   &        \\
\object{19118+0945}   & L   & 19 14 29.7 $\pm$ 1.6    &     09 51 47  $\pm$  46  & 19 14 14  &      09 50 33  &      &  246  &   61.25   &      \\
\object{19410+2336}   & S   & 19 43 12.2 $\pm$ 1.1    &     23 44 03  $\pm$  12  & 19 43 11  &      23 44 20  & 2.1  &   21  &   20.67   &       \\
      \midrule
      \multicolumn{9}{l}{\small\sl Sources with OH masers but without millimeter emission}\\
      \cmidrule(l){1-2}
\object{19217+1651} & \object{19374+2352} & \object{19388+2357} & \object{20227+4154} & & & & & &  \\
%\object{19217+1651}   &  S   & 19 23 57.9 $\pm$ 0.9    &     16 56 42  $\pm$  13  &   0.21   & $-$2.0   &  10.0    &   1.35 & 0.05 & 1  &  \\
%\object{19374+2352}   &  H   & 19 39 37.4 $\pm$ 4.0    &     23 59 53  $\pm$ 109  &  37.06   &   35.0   &  40.0    &   0.46 & 0.06 &    &  \\
%\object{19388+2357}   &  H   & 19 41 10.2 $\pm$ 2.2    &     24 03 44  $\pm$  25  &  35.81   &   34.0   &  39.0    &   0.32 & 0.04 & 77 &  \\
%\object{20227+4154}   &  H   & 20 24 34.6 $\pm$ 1.3    &     42 06 12  $\pm$  22  &  24.29   &   10.0   &  25.0    &   0.80 & 0.07 &    &  \\

 \bottomrule
%%%%%%%%%%%%%%%%%%%%%%%%%%%%%%%%%%%%%%%%%%%%%%%%%%%%%%%%%%%%%%%%%%%%%%%%%%%%%%%%%%%%%%%%%%%%%%%%%%%%%%%%%%%%%%%%%%%%%%%%%%%%
    \end{supertabular}
 \label{tab:OHmaserRes}
  \end{center}
}
%\end{landscape}
\twocolumn

\subsection{Comment on individual sources}

\begin{figure*}
%%========================================================================%%
  \resizebox{\hsize}{!}{
  \includegraphics[angle=-90,width=6cm]{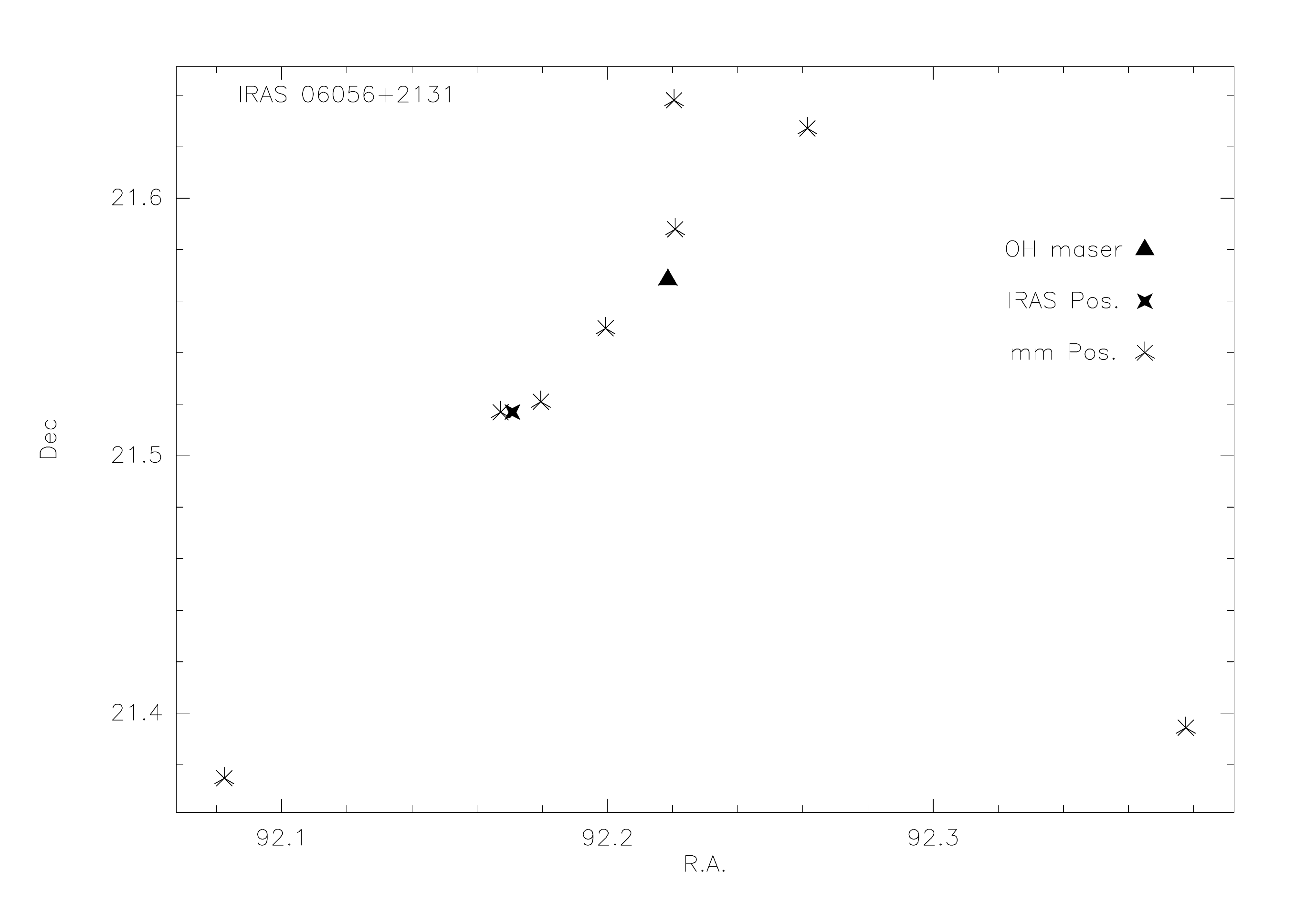}
  \includegraphics[angle=-90,width=6cm]{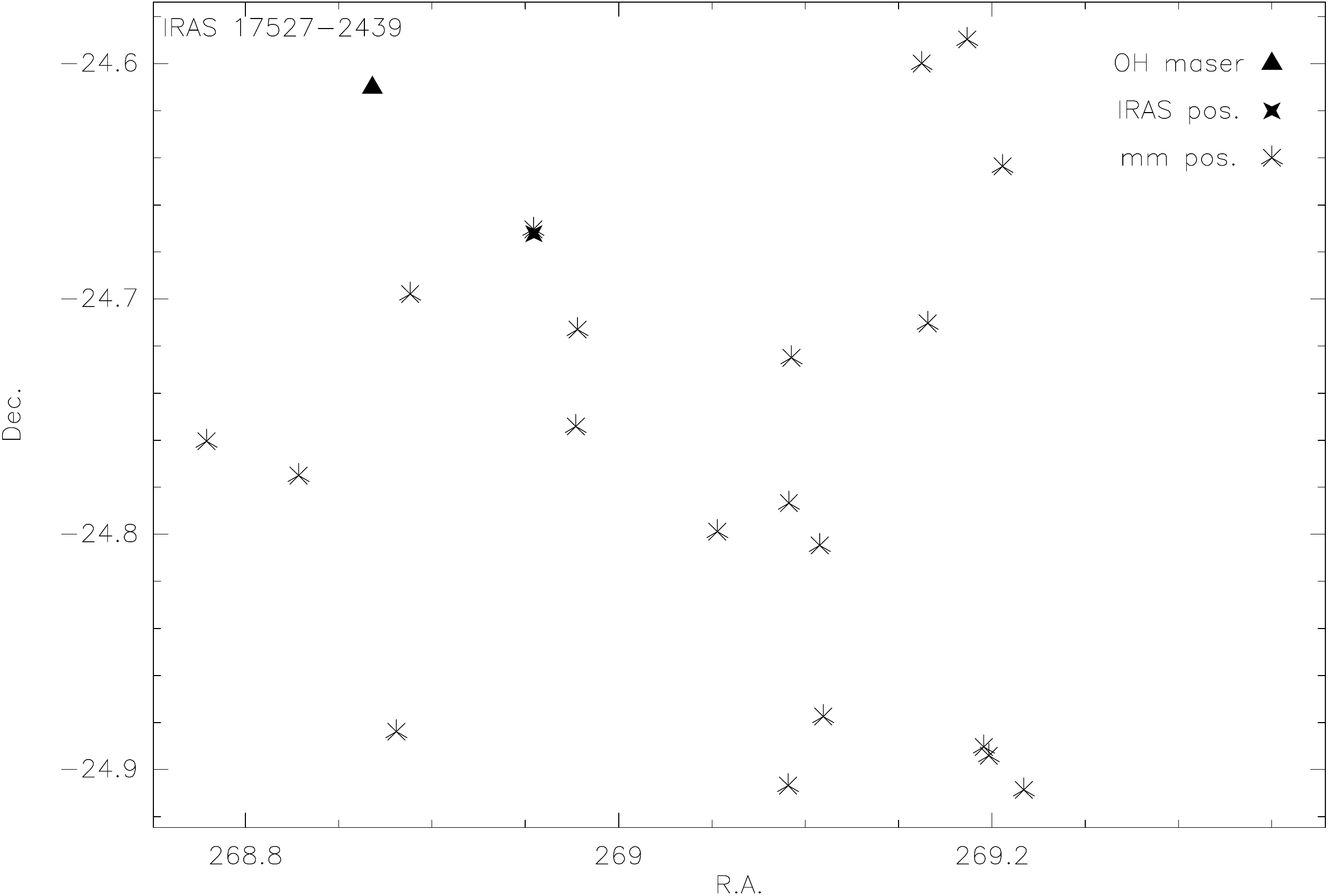}
}
%%========================================================================%%
  \resizebox{\hsize}{!}{
  \includegraphics[angle=-90,width=6cm]{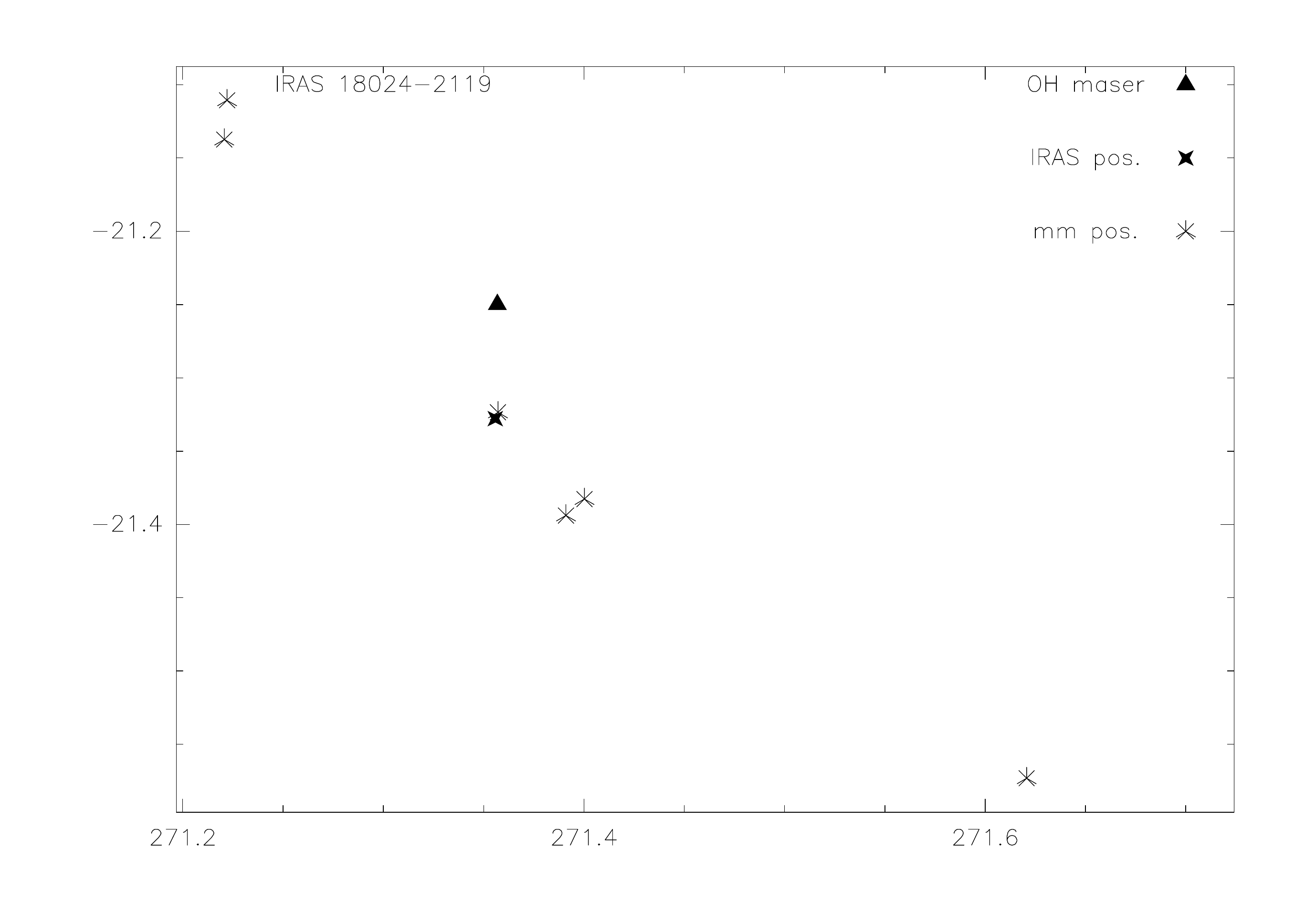}
  \includegraphics[angle=-90,width=6cm]{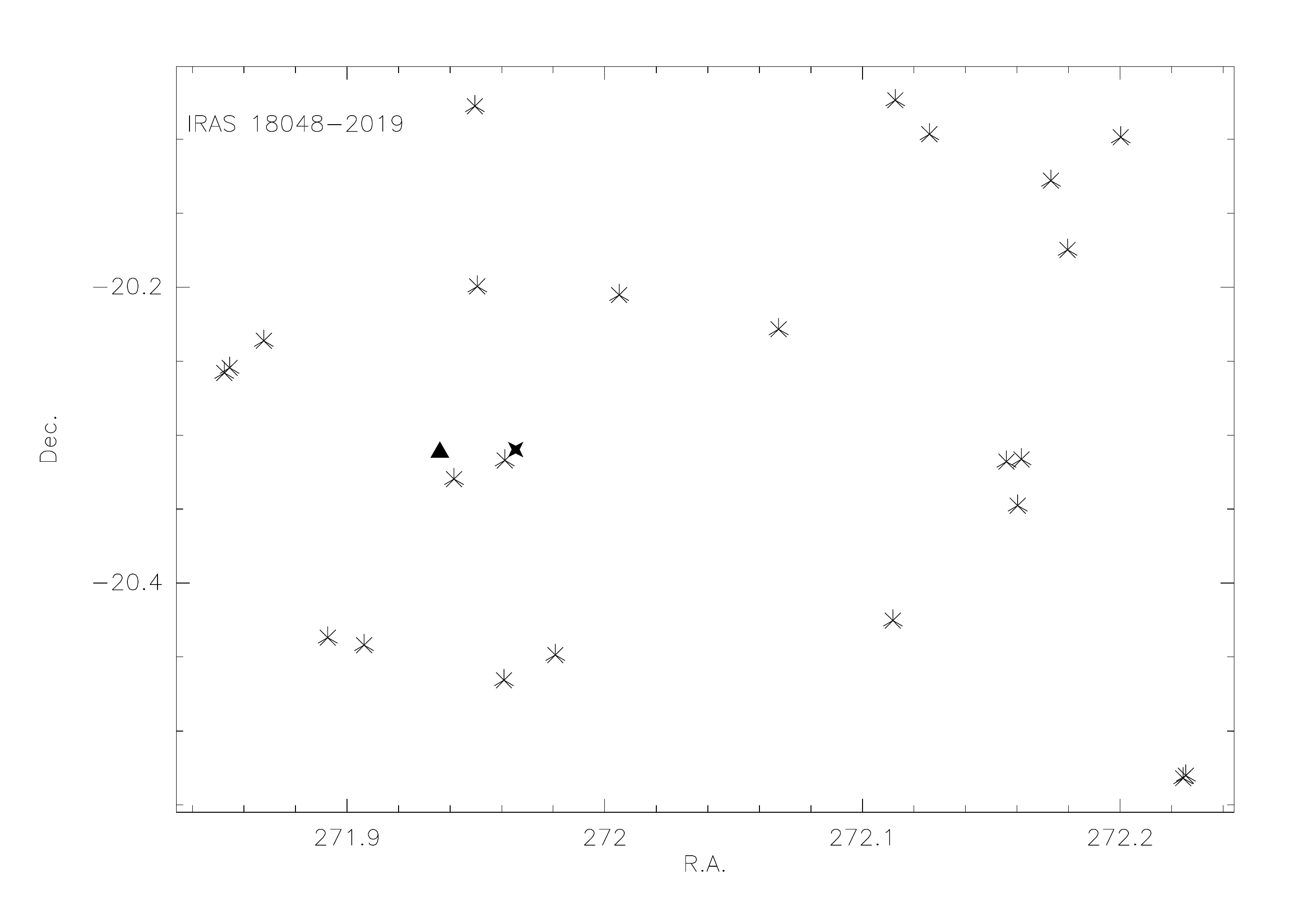}
}
%%========================================================================%%
  \resizebox{\hsize}{!}{
  \includegraphics[angle=-90,width=6cm]{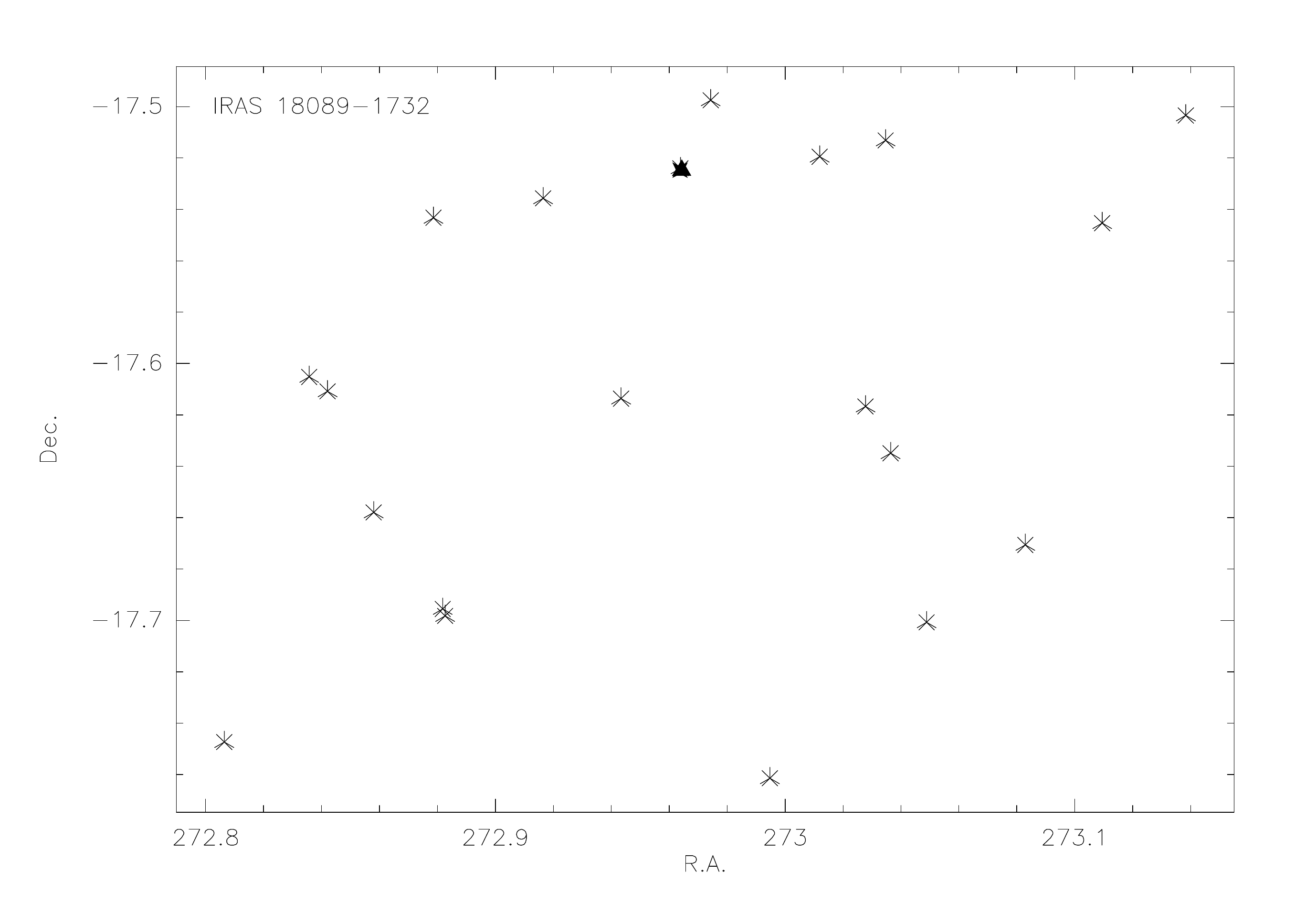}
  \includegraphics[angle=-90,width=6cm]{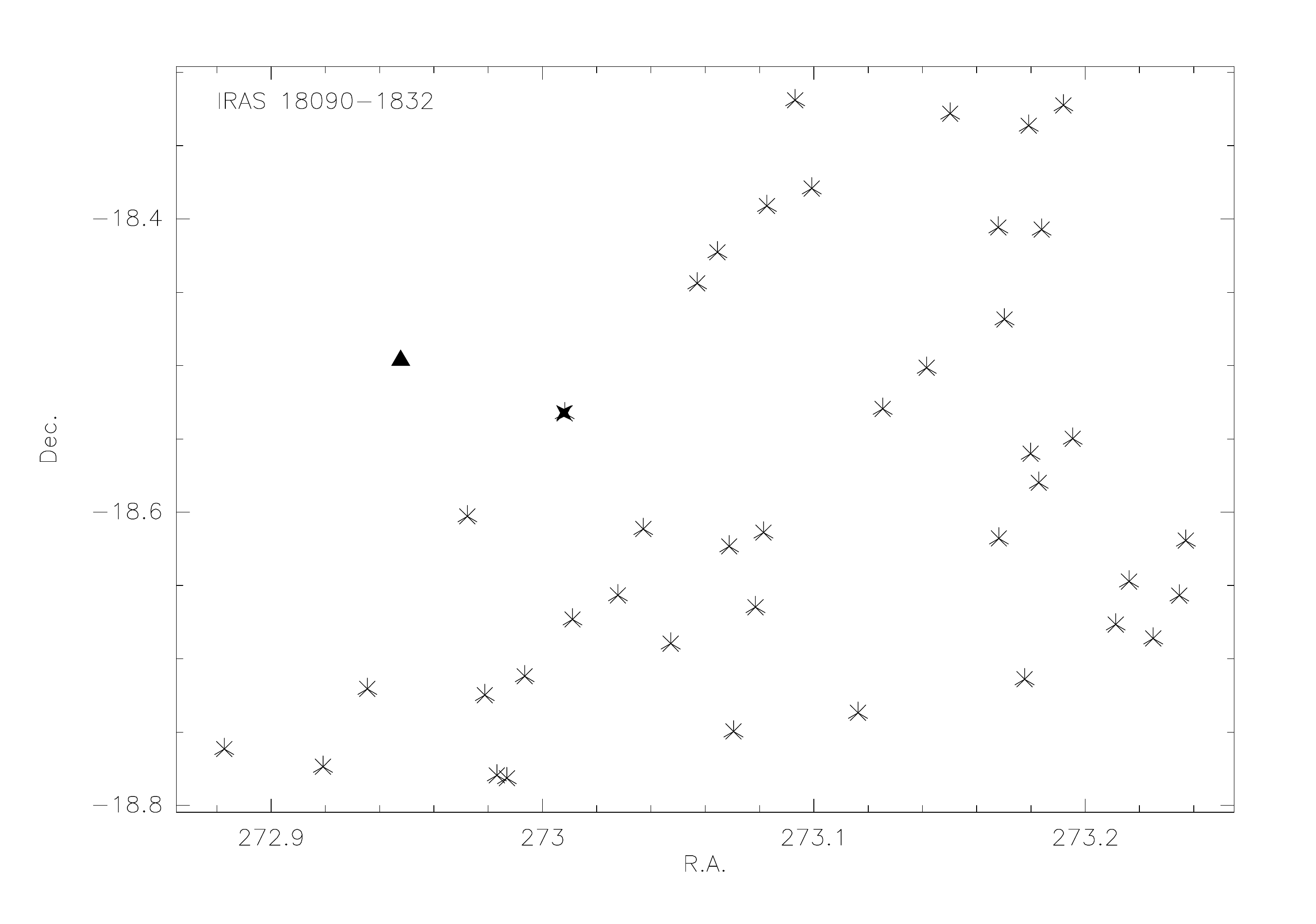}
}
%%========================================================================%%
  \resizebox{\hsize}{!}{
  \includegraphics[angle=-90,width=5cm]{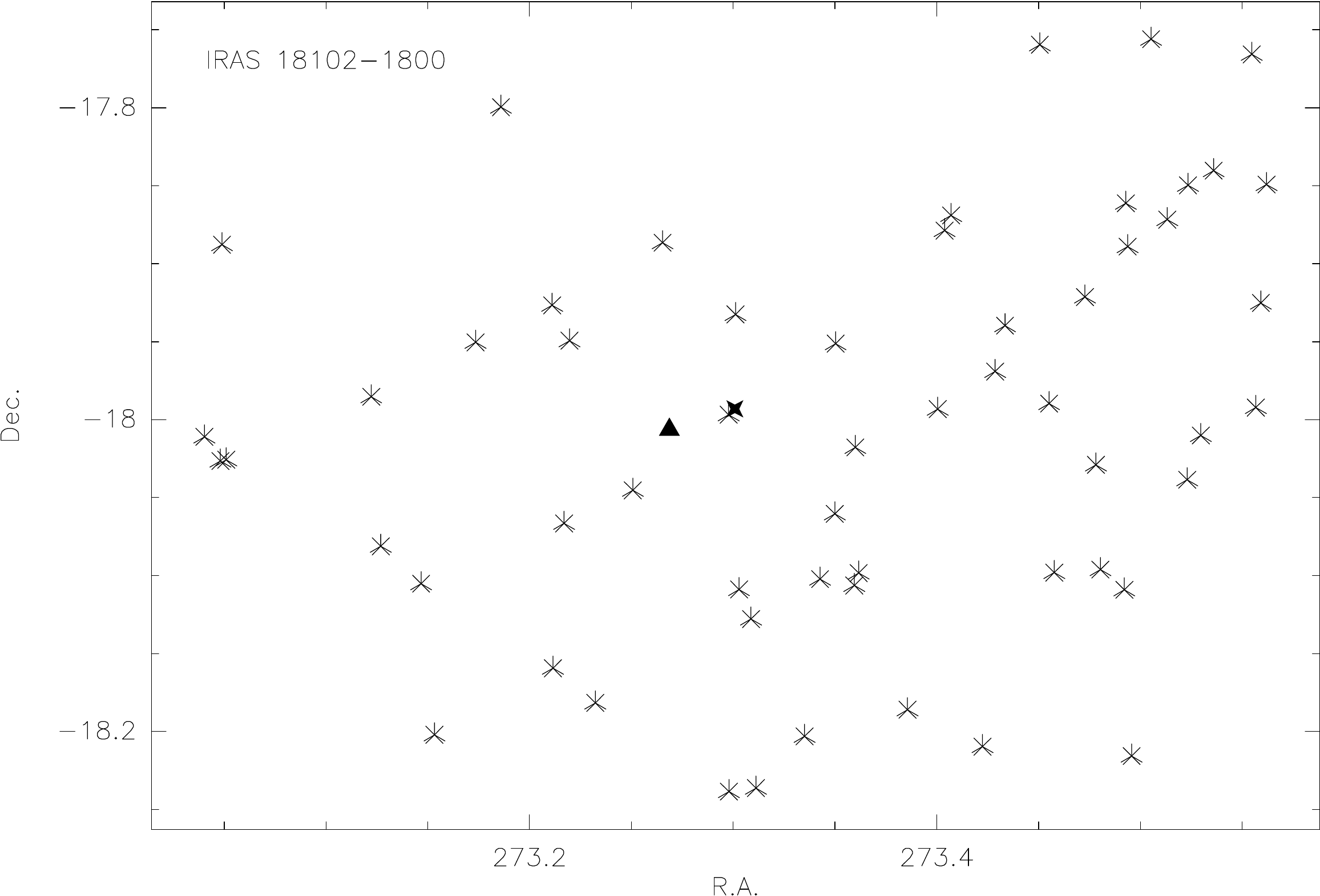}
  \includegraphics[angle=-90,width=5cm]{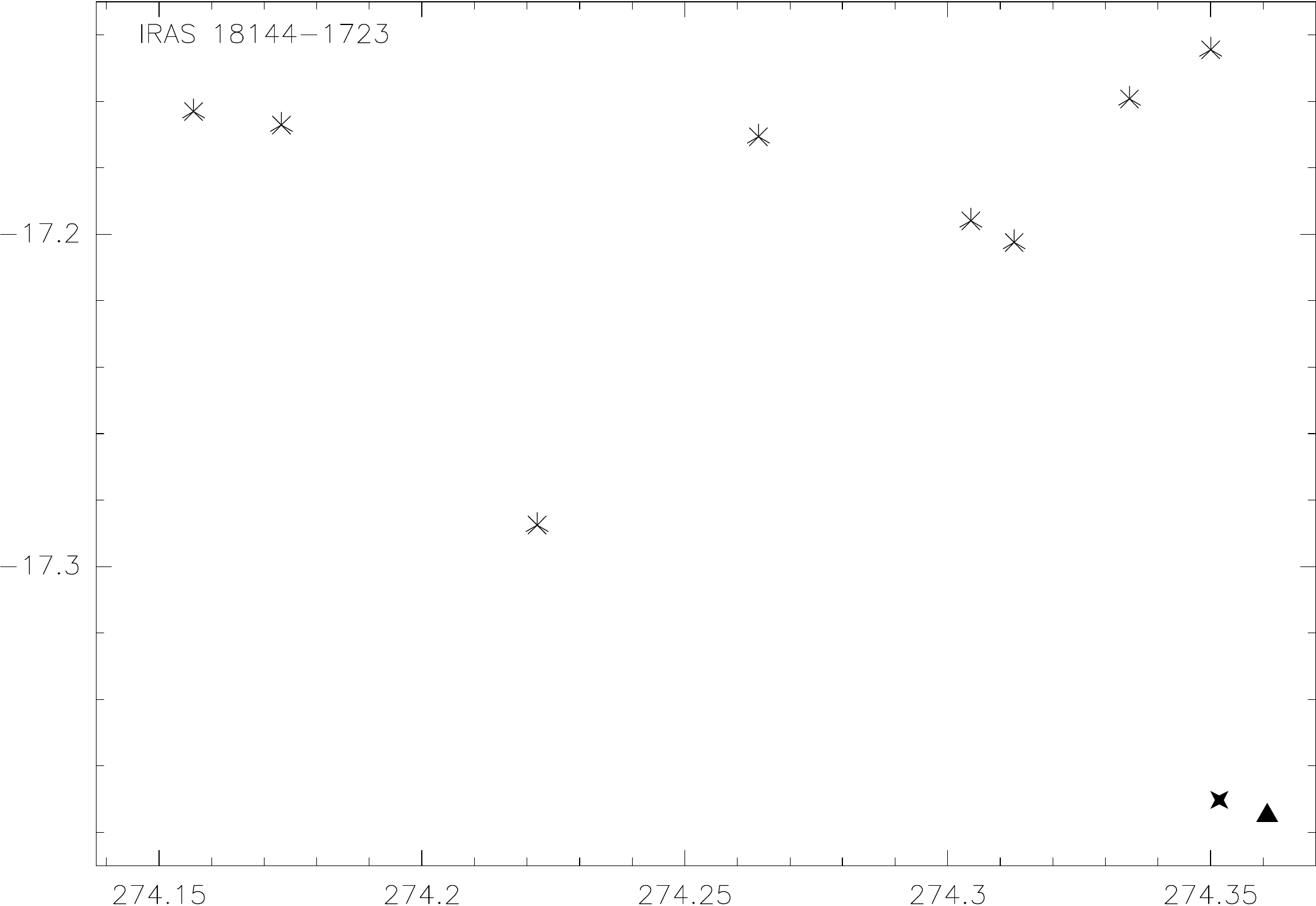}
}
%%========================================================================%%
\caption{The position of the OH masers and all mm peaks found within $\sim$ 30 arcmin as well as the IRAS position. The IRAS name of the source is indicated in the upper left corner of the first three plots. The x-axis represents the right ascension and the y-axis represents the   declination, both in degrees of arc.}
\label{fig:1mm}
  \end{figure*}

\textbf{IRAS 06056+2131}.
The OH masers were detected towards this source at 1665-, 1667-, and 1720-MHz lines. The velocity of the 1720-MHz OH satellite line is more consistent with the NH$_3$ gas velocity (Molinari et al. \citealp{molinari96}) than the two OH main lines.
The CO outflow map of the region presented by Zhang et al. (\citealp{zhang05}) show at least two outflows. The IRAS source is consistent with one of these outflows and the OH main lines may be consistent with the other. Several 1.1-mm peaks were detected towards this region mostly in the shape of bow shock (Figure \ref{fig:1mm}). The peak seems to be closer to the OH main lines is much weaker than the two peaks closer to the IRAS source. The velocity of the \methanol\ maser source detected by Szymczak et al. (\citealp{szymczak00}) is consistent with the velocities of the OH maser main lines ($\sim$ 10 km \s). All these indicate that this region is complex and contain at least two protostars in different evolutionary stages. This is consistent with the near-IR K$_s$ band observations of Faustini et al (\citealp{faustini09}) which suggest a cluster of several members with a radius of 0.3 pc.

\textbf{IRAS 17527$-$2439}.
 The 1665-MHz OH maser detected towards this region at velocity of 11.5 km \s is more consistent with the NH$_3$ gas velocity (13.2 km \s) than the \htwoo\ maser detected at velocity of $\sim$ -2 km \s by palla et al. (\citealp{palla91}). The 1.1-mm plot (Figure \ref{fig:1mm}) show that one of the peaks is coincident with the IRAS position while the OH maser is offset by $\sim$ 5 arc-minute.

\textbf{IRAS 18024$-$2119}.
The OH masers is offset by $\sim$ 4 arc-minutes from the 1.1-mm peak which is more consistent with the IRAS position and the 850 $mu$m continuum (Molinari et al. \citealp{molinari08}). No cluster was detected by Faustini et al (\citealp{faustini09}).

\textbf{IRAS 18048$-$2019}.
The OH masers (detected at the 1665-, 1667-MHz, and 1720-MHz) and IRAS source are associated with different 1.1-mm peaks and the one closer to the IRAS source is $\sim$ 7 times stronger than the one closer to the OH maser. This source is also associated with much stronger 6.7-GHz \methanol\ and \htwoo\ masers (Schutte et al. \citealp{Schutte93}; Palla et al \citealp{palla91}) peaks at different velocities but at similar velocity ranges. This source shows absorption feature at the 1667-MHz OH and a weak maser at the 1720-MHz satellite line which indicates that this source may be associated with SNR.

\textbf{IRAS 18089$-$1732 (G12.89+0.49)}.
The OH maser towards this source was firstly detected by Cohen et al. (\citealp{cohen88}) at the 1665-MHz line. Argon et al. (\citealp{argon00}) mapped this line in arc-second resolution which show three different components. The component which associated with strongest emission ($\sim$ 4 Jy) is consistent with the IRAS position and the others are offset by $\sim$ 1 and 2.7 arcsec. EFC07 detected the 1665-MHz line as well, but with ten times stronger flux density ($\sim$ 30 Jy). EFC07 also detected an emission at the 1667-MHz line. This means that this is a variable source which is consistent with the daily monitoring nine-year observations of Goedhart et al. (\citealp{goedhart09}) at the 6.7 and 12.2 GHz \methanol\ maser lines. They measure a period of variability as short as less than a month. The OH masers, 1.1-mm peak, and the IRAS position may be in coincident (Figure \ref{fig:1mm}). These source is associated with \htwoo\ (palla et al. \citealp{palla91}) and \methanol\ (Szymczak et al. \citealp{szymczak00}) masers and is also associated with very weak 3.6 cm continuum emission, 0.9 mJy (Sridharan et al. \citealp{sridharan02}). Recent high angular resolution submillimeter observations in various spectral lines by Beuther et al. (\citealp{beuther05}) detect a massive rotating structure perpendicular to an emanating outflow which is likely associated with the central accretion disk.

\textbf{IRAS 18090$-$1832}.
The OH maser emission detected towards this source at the two main lines (RHC only) is much weaker ($\sim$ 0.8 Jy) than a relatively strong, 77 Jy 6.7-GHz \methanol\ maser (Szymczak et al. \citealp{szymczak00}). The peak velocity of the \methanol\ maser is more consistent with the 1667-MHz line than the 1665-MHz. One of the mm peaks is coincident with the IRAS source while the OH maser is offset by $\sim$ 4 arcmin without any closer mm emission peak.

\textbf{IRAS 18102$-$1800}.
The OH maser detected towards this sources at the 1665-MHz (RHC) only. One of the mm peaks is more consistent with the IRAS source than the OH maser which is offset by $\sim$ 3 arcmin. This source is associated with \methanol\ maser and a 44 mJy radio emission (Sridharan et al. \citealp{sridharan02} and reference therein).

\textbf{IRAS 18144$-$1723}.
The OH maser detected towards this sources at the main lines is a relatively stronger (80 Jy) than the other types of masers, \methanol\, $\sim$ 33 Jy (Szymczak et al. \citealp{szymczak00}) and H2O, $\sim$ 24 Jy (Palla et al. \citealp{palla91}). There is a significant gap, 13 km \s\ between the central velocities of strongest components in the 1665 and 1667 MHz lines. The OH maser and the IRAS source are not closely associated and are not consistent with any of the mm peaks to within $\sim$ 18 arcmin. The radio continuum emission detected towards this region at 2 and 6 cm is offset from the IRAS source by $\sim$ 90 arcsec (Molinari et al. \citealp{molinari98}) and no outflow was detected by Zhang et al. (\citealp{zhang05}).

\begin{figure}
  \centering
  \includegraphics[angle=-90,width=9cm]{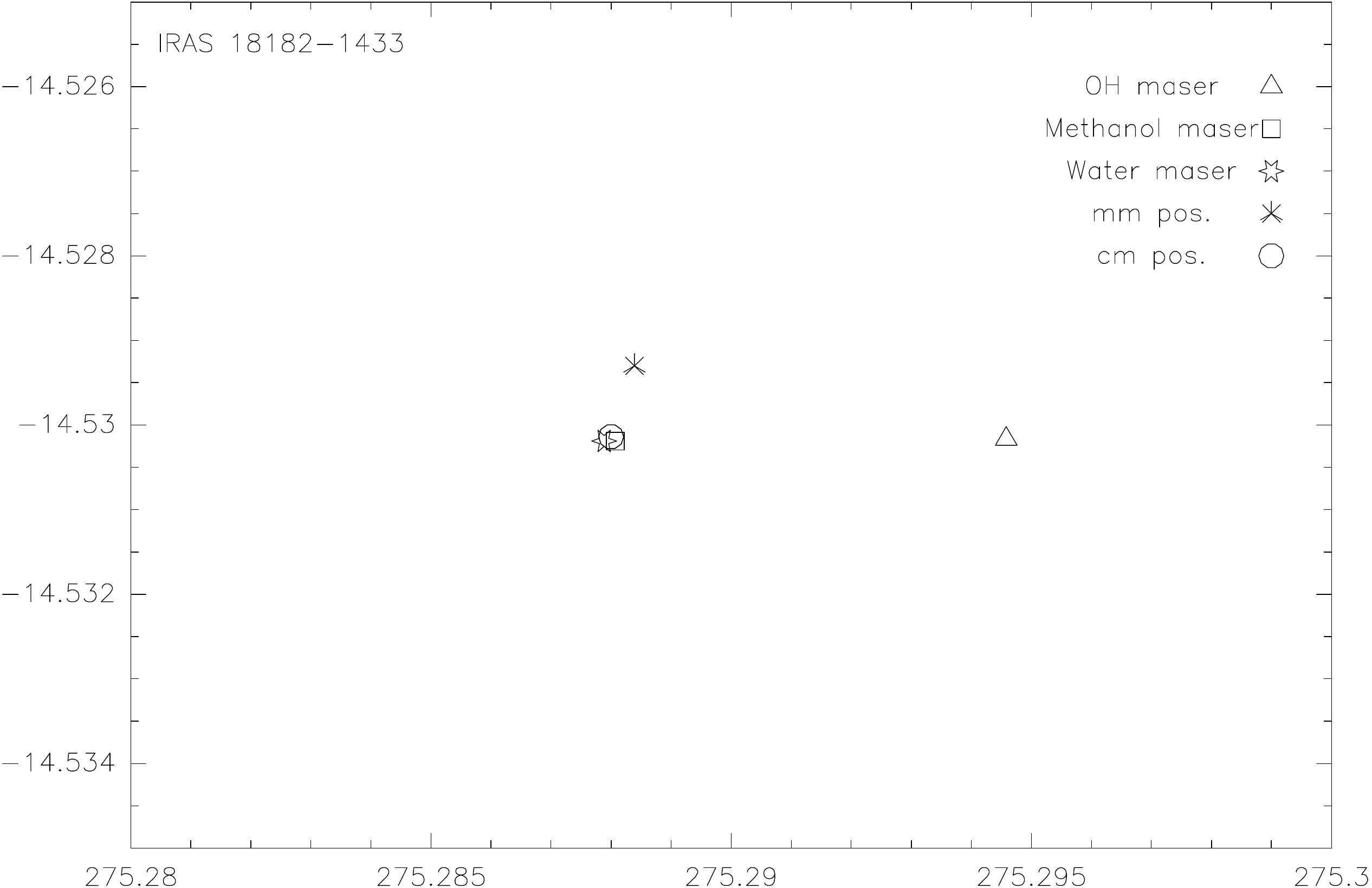}
\caption{The OH (from Foster \& Caswell 1999), \methanol\, and \htwoo\ masers (from Zapata el al. \citealp{zapata06}, cm and the closest mm peak in the region of IRAS 18182$-$1433. The x-axis represents the right ascension and the y-axis represents the declination, both in degrees of arc}
\label{fig:18182}
\end{figure}

\textbf{IRAS 18182$-$1433}.
The velocity of OH maser detected towards this source at the two main lines is consistent with the \methanol\ and \htwoo\ masers detected by Szymczak et al. (\citealp{szymczak00}) and Beuther et al. (\citealp{beuther02a}) respectively. But although these two later traces were mapped recently using the VLBI by Sanna et al. (2010) may be coincident with VLA 3.6 cm observations by Zapata el al. (\citealp{zapata06}), the OH maser mapped by Forster \& Caswell (1999) is offset by $\sim$ 25 arcsec and the closest mm peak is offset by $\sim$ 3 arcsec (figure \ref{fig:18182}). An outflows was detected by Beuther et al. \cite{beuther02a}.

\textbf{IRAS 18566+0408}.
Towards this source one of the 1.1 mm peaks is consistent with the IRAS positions. This has also been found with 1.2 mm observations by Beuther et al. \cite{beuther02a}. The OH masers have a wide velocity range ($\sim$~40 km~\s) and the two positions detected by EFC07 at the two mainlines are located in the two opposite sides of the IRAS/mm source (figure \ref{fig:1mm}). This mm peak has the strongest flux among all other peaks within 30 arcmin. This source is also associated with \methanol\ maser emission (Sridharan et al. \citealp{sridharan02}; Szymczak et al. \citealp{szymczak00}). The \htwoo\ maser emission was detected by Sridharan et al. (\citealp{sridharan02}), while was not detected by Palla et al. (1991). An outflow was detected by Beuther et al. (\citealp{beuther02a}) and Sridharan et al. (\citealp{sridharan02}) place an upper limit of 1 mJy on the 3.6 cm radio continuum flux from any source in this region.

\begin{figure*}
%%========================================================================%%
  \resizebox{\hsize}{!}{
  \includegraphics[angle=-90,width=7cm]{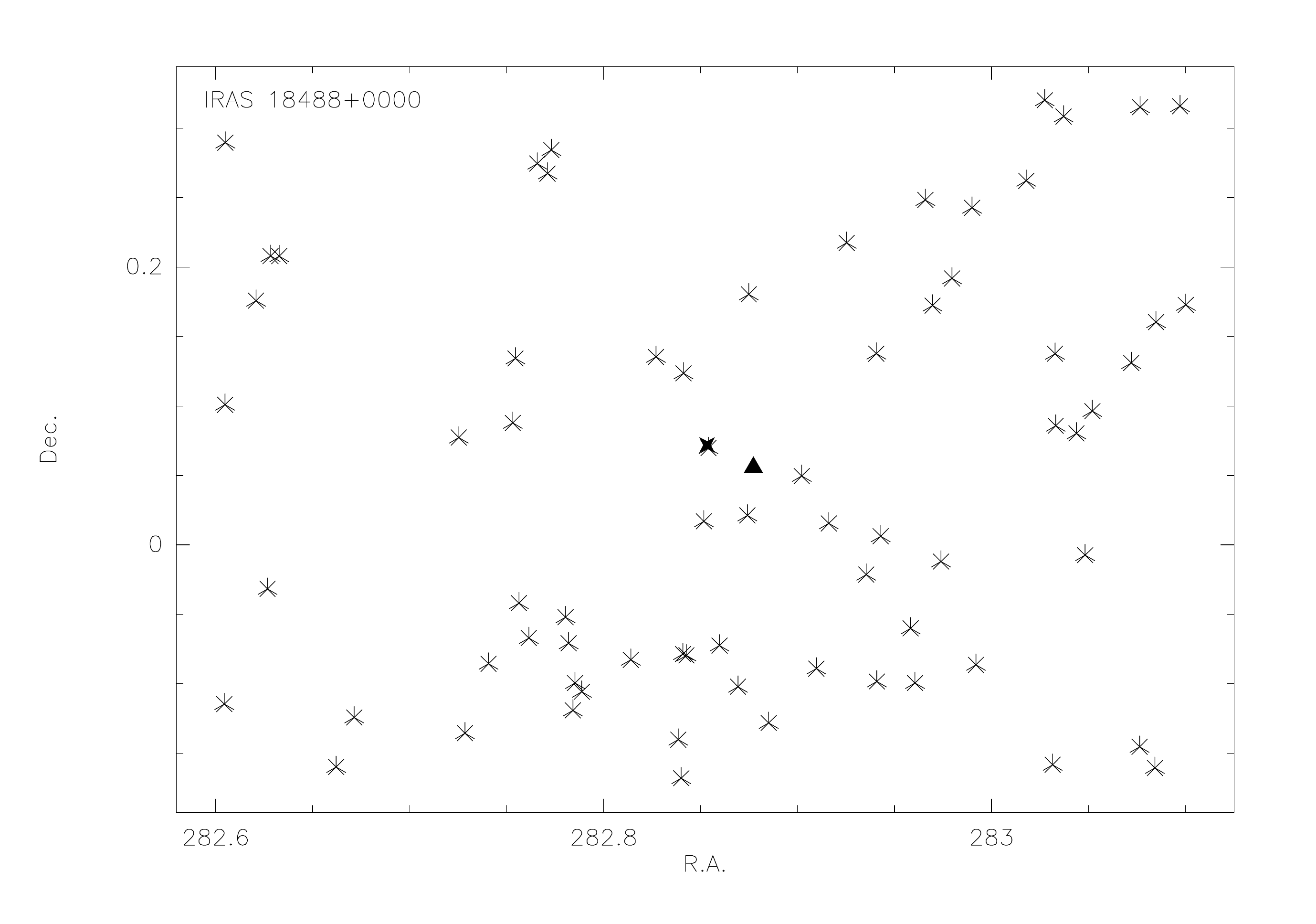}
  \includegraphics[angle=-90,width=7cm]{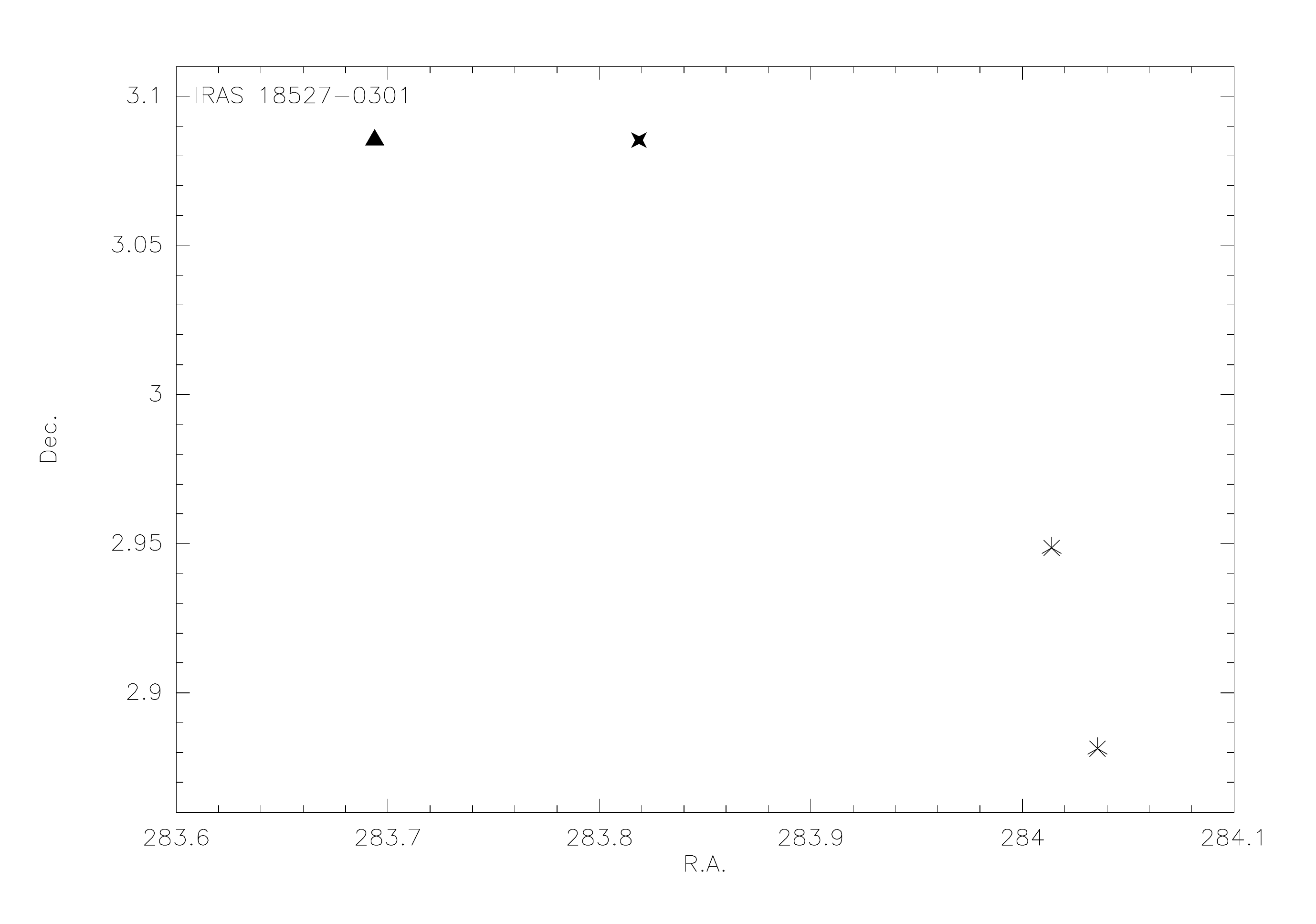}
}
%%========================================================================%%
  \resizebox{\hsize}{!}{
  \includegraphics[angle=-90,width=7cm]{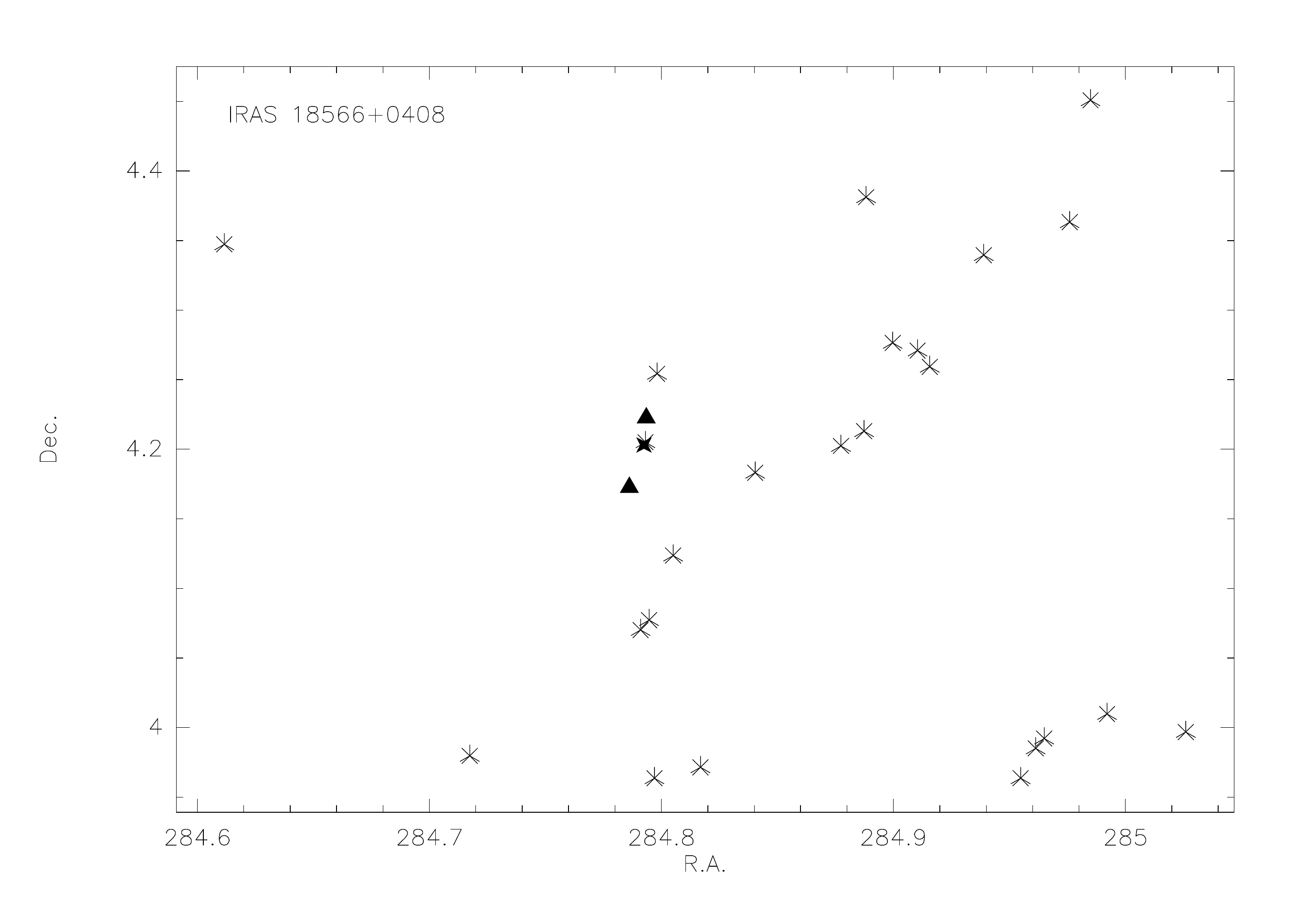}
  \includegraphics[angle=-90,width=7cm]{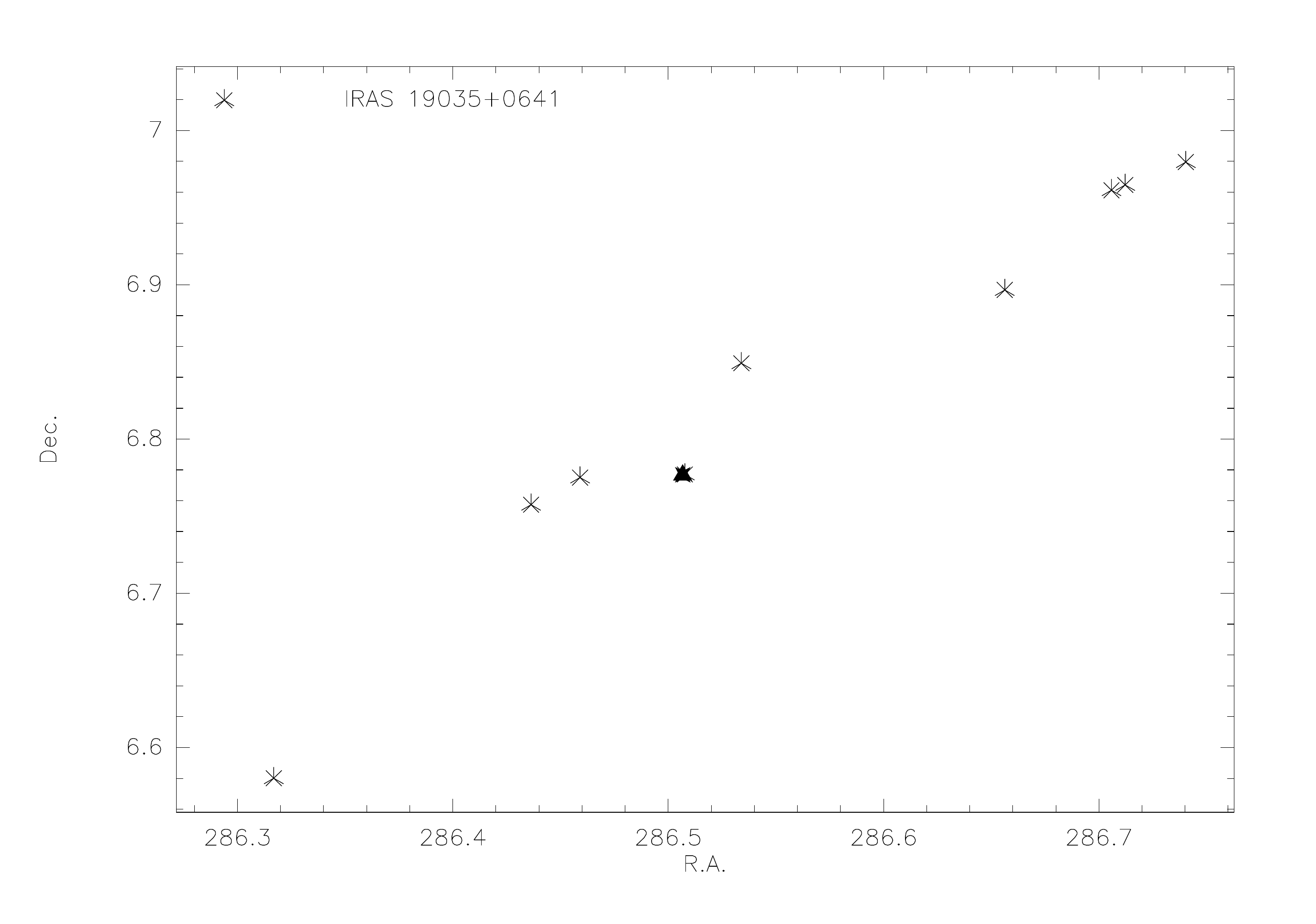}
}
%%========================================================================%%
  \resizebox{\hsize}{!}{
  \includegraphics[angle=-90,width=7cm]{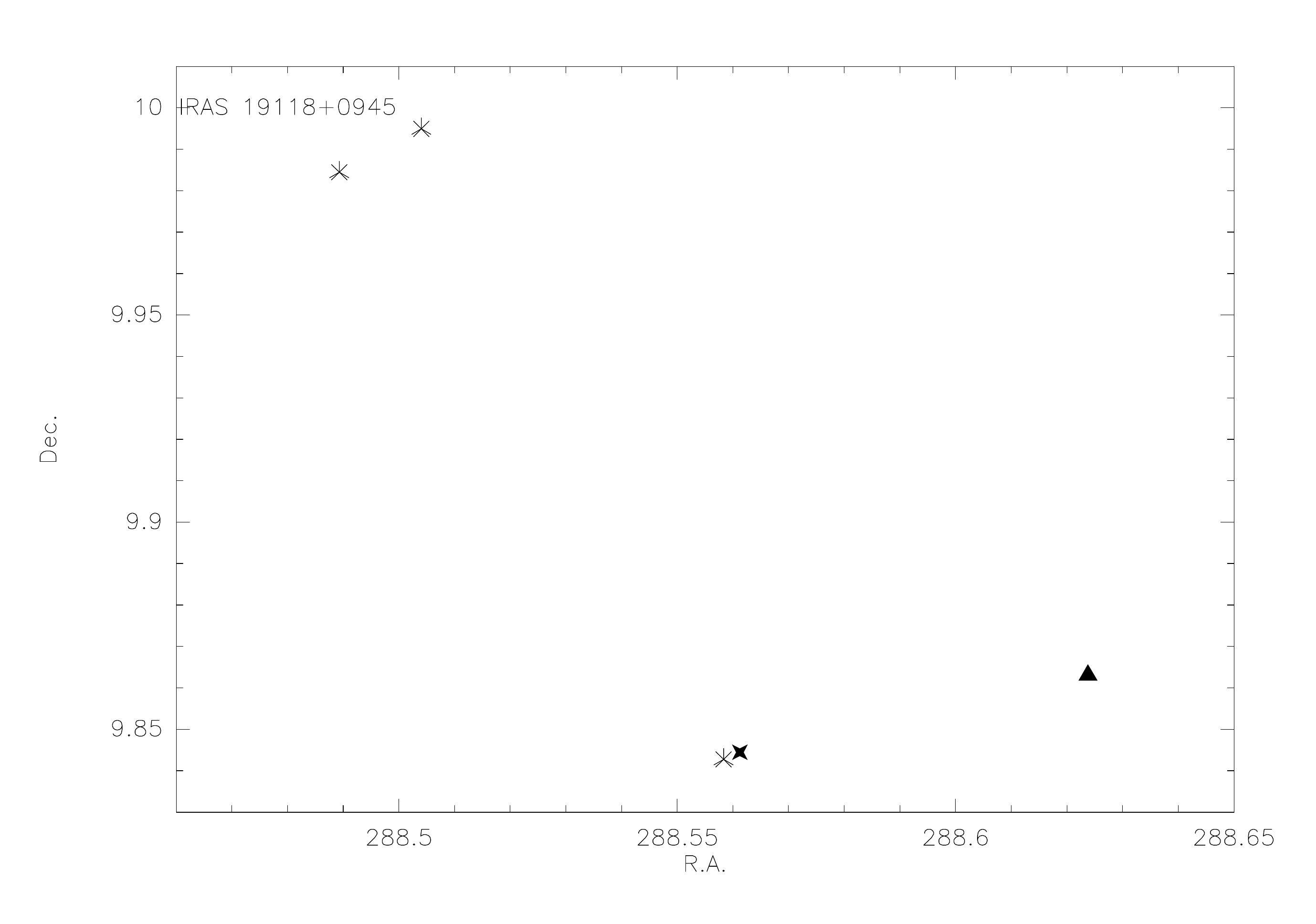}
  \includegraphics[angle=-90,width=7cm]{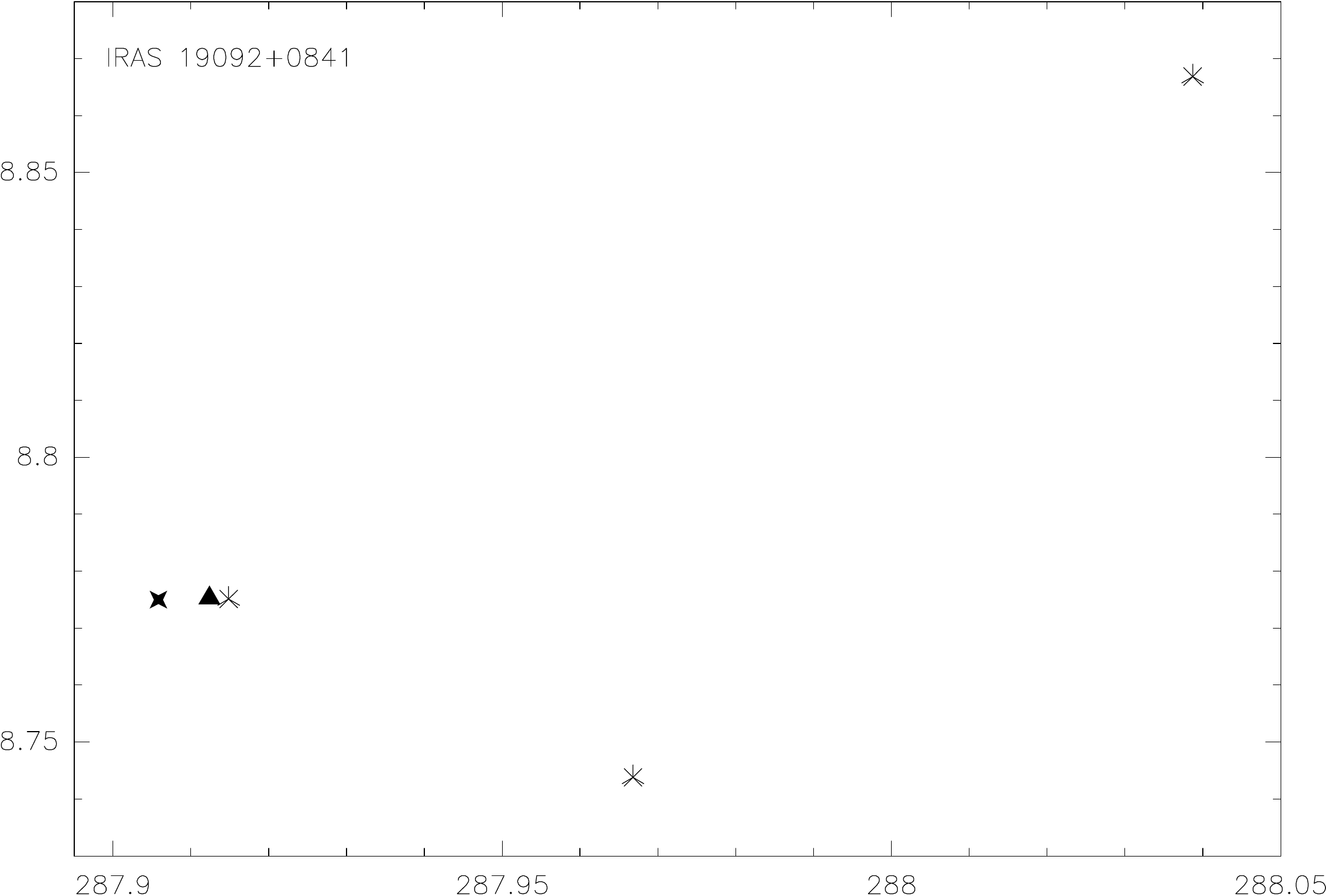}
}
%%========================================================================%%
%  \resizebox{\hsize}{!}{
\begin{center}
  \includegraphics[angle=-90,width=7cm]{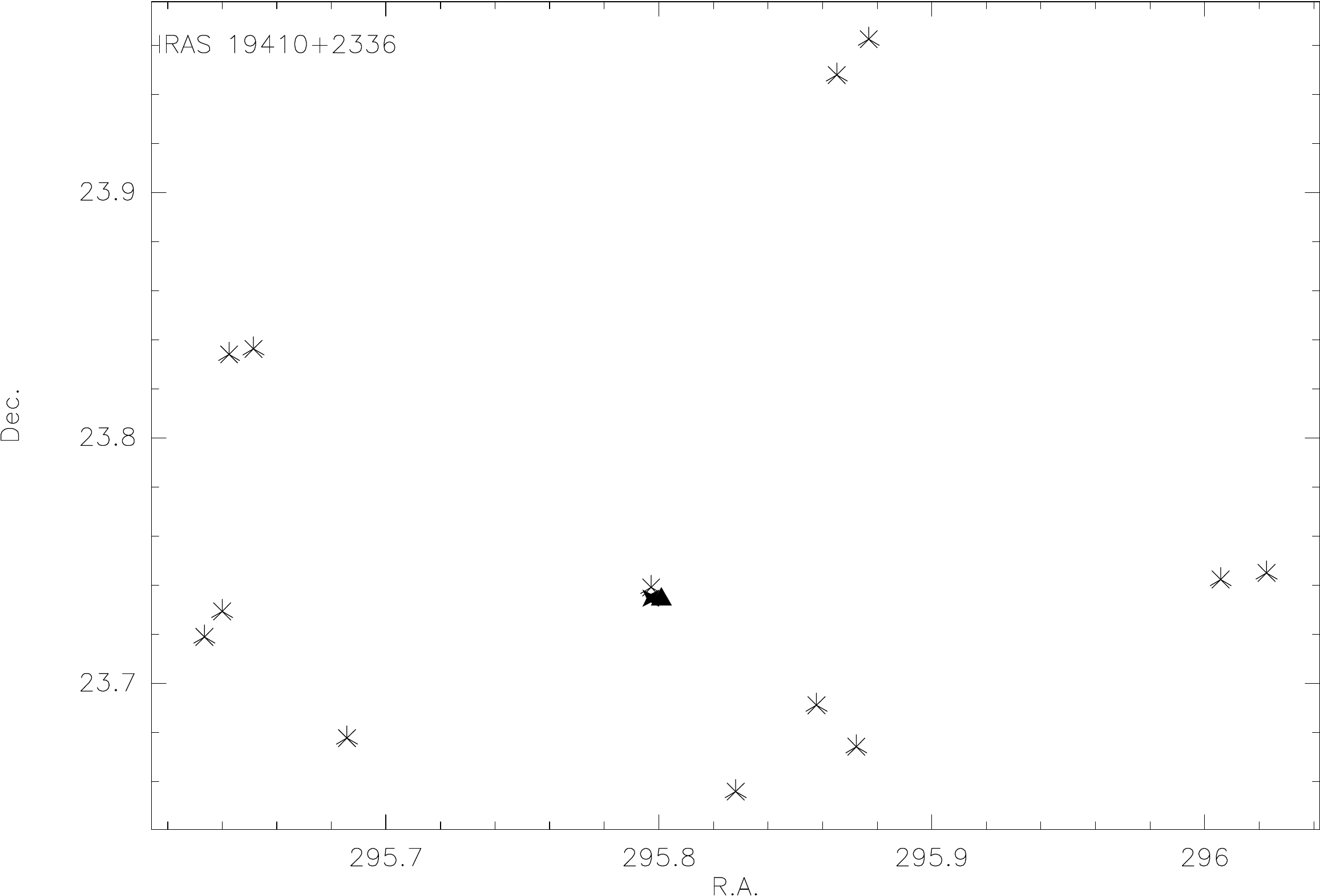}
%}
  \end{center}
%%========================================================================%%
\caption{same as Figure (\ref{fig:1mm}).}
\label{fig:2mm}
  \end{figure*}

\textbf{IRAS 19035+0841 (G40.622-0.137)}.
This is the second source with the three tracers, OH, mm, and IRAS are in very good agreement to within 0.05 pc (figure~\ref{fig:2mm}). The position of the OH maser is taken from the arcsec resolution observations of Argon et al. (2000). It is the strongest OH maser source ($\sim$~300 Jy) within this sample. The mm peak consist with the OH maser has the strongest flux of the other peaks within 30~arcsec. An outflow and 1.2 mm continuum emission were detected by Beuther et al. (\citealp{beuther02a}).

\textbf{IRAS 19092+0841}.
The OH maser position measured by high resolution observations using MERLIN (Edris et al. 2009) is 0.2 pc offset from the mm peak. It seems to be more closer than the IRAS source. A VLA observations of 44-GHz class I methanol masers (Kurtz et al. 2004) show a close association between this tracer and the OH masers. This source is also associated with class II \methanol\ and \htwoo\ masers.

\textbf{IRAS 19118+0945}.
The mm and IRAS are seemingly closer (to within 12 aresec) than the OH maser which is offset by $\sim$~4 arcmin. No \htwoo\ or \methanol\ masers have been detected towards this source (Palla et al. \citealp{palla91}; Szymczak et al. \citealp{szymczak00}).

\textbf{IRAS 19410+2336 (G59.78+0.06)}.
The IRAS, OH and 1.1-mm emission are closely associated within 0.2 pc and consistent with previous 1.2-mm continuum observations as well as \methanol\ maser(Minier et al. \citealp{minier05}). This observations show another weaker mm clump with no mid-IR emission. It is associated with weak radio source of $\sim$ 1 mJy (Sridharan et al. \citealp{sridharan02}). The OH maser associated with this source at only 1665-MHz line has one feature at velocity of $\sim$ 20.6 km \s\ which falling in the \methanol\ masers velocity range (14 to 28 km \s) with several features peak at 17 km \s\ (Szymczak et al. 2000). Two outflows was detected by Beuther et al. \cite{beuther02a} to associate each of the 1.2 mm clumps.

\begin{figure}
  \centering
  \includegraphics[angle=-90,width=9cm]{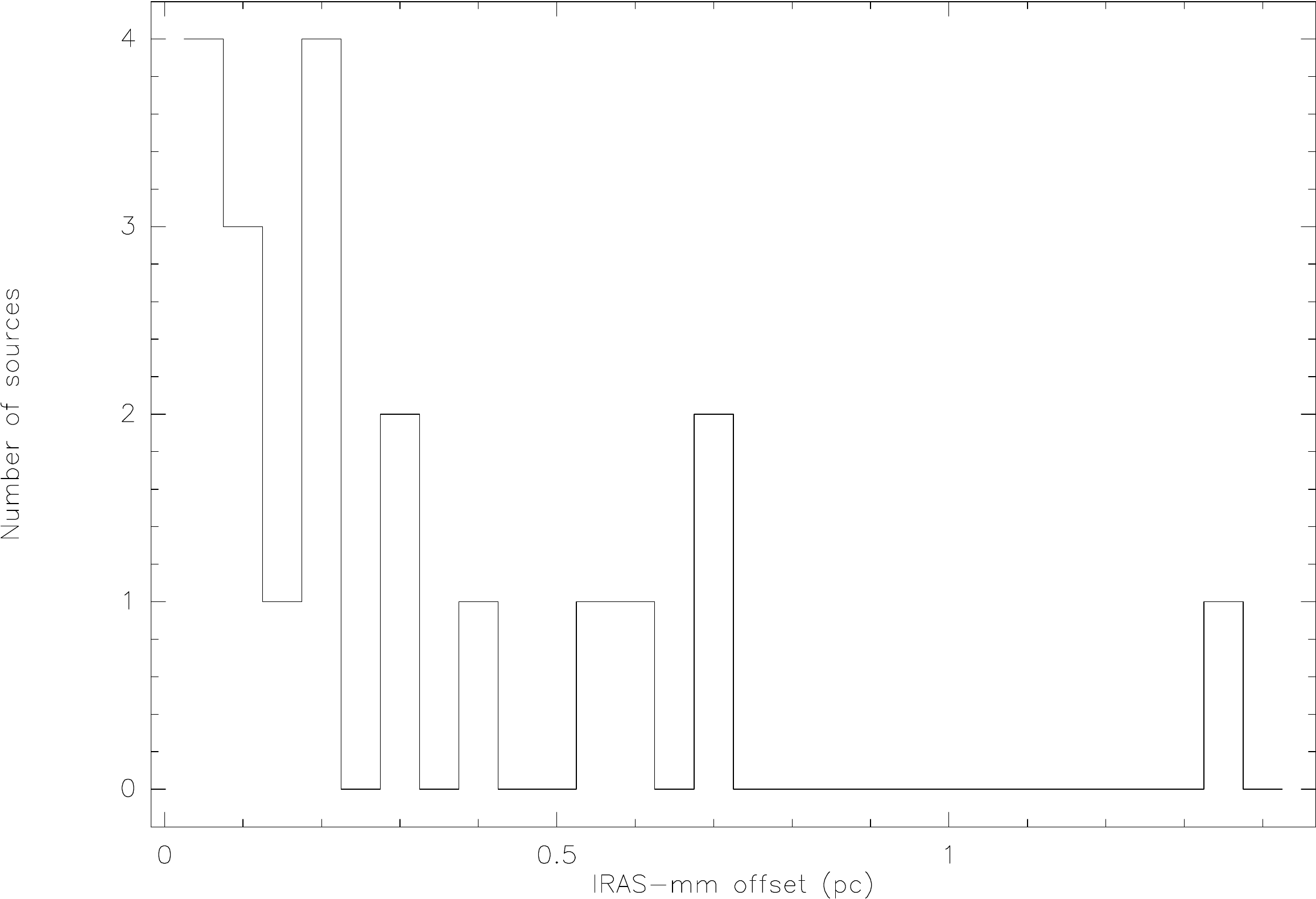}
\caption{The offsets of the IRAS sources and the closest mm peak. All IRAS sources are close to the mm peaks to $<$ 1.4 except for two sources, IRAS 18144-1723 and IRAS 18527+0301 which are not plotted here.}
\label{fig:offset}
\end{figure}

\section{Discussion}

It obvious that for those sources with high angular resolution observations, their are a very good agreement between the OH masers and the millimeter emission as well as the IRAS position. Indeed towards IRAS $18089$-$1732$ and IRAS $19035+0641$, the three tracers seem to come from the same core. This indicate that high angular resolution observations could improve this consistency. This should not remove the possibility that towards some of these sources the OH masers are associate but not consist with the millimeter emission. These sources could be in very early evolutionary stage with the OH masers coming from the external shell of the collapsing core. One of these sources is IRAS $19092+0841$ for which the OH masers have been observed in high angular resolution using MERLIN (Edris et al. 2009). The OH maser is offset from the millimeter peak by $\sim$ 9 arcsec.

\subsection{The association of OH masers and dust emissions}

The association of infrared emission and maser emission has been confirmed by many surveys. Many surveys of star forming regions has used the IRAS catalogue via color-selected sources (e.g. Palla et. al. 1991). The detection rate of up to 26 \% of the maser emission has been found towards these IR sources. Also The association between the maser emission and mm and sub-mm have been found in the case of methanol masers (e.g. Breen et al. 2010) and water masers (e.g. Jenness et al. 1995).the association OH masers and the mm emission has been found towards some sources (e.g. Edris et al. 2005).
Towards This sample OH maser emission is in more consistent with the mm peaks than the IR peaks. This means that the OH maser is associated with the colder dust more than the hotter one. This consistent with the maser being originated from the outflows or the outer layers of the circumstellar material. The Elitzur \& de Jong (1978) model propose that the OH masers originated from the outer shell of the HII region. however there is no detectable HII regions towards these sources. Gray et al. \citealp{gray03} propose that towards some sources the OH masers are originated from different layers of the circumstellar disk. The criteria of choosing these sample would prefer the later model but high angular resolution observations are needed to confirm this suggestion.

\section{Conclusions}

The association between OH maser emission and 1.1 mm continuum emission towards 27 IRAS sources has been studied. The OH maser emission is more consistent with the mm peaks than the IR peaks. The large positional uncertainty do not allow to give a firm conclusions but towards the sources which have been observed at high angular resolutions the three tracers seem to be in correlation. High angular resolution observations are needed to reach to a conclusion about the association between the maser emission and the dust emission.
%\vspace{1cm}

%ACKNOWLEDGMENTS

%%========================================================================%%
\begin{figure*}
%%========================================================================%%
  \resizebox{\hsize}{!}{
  \includegraphics[angle=-90,width=7cm]{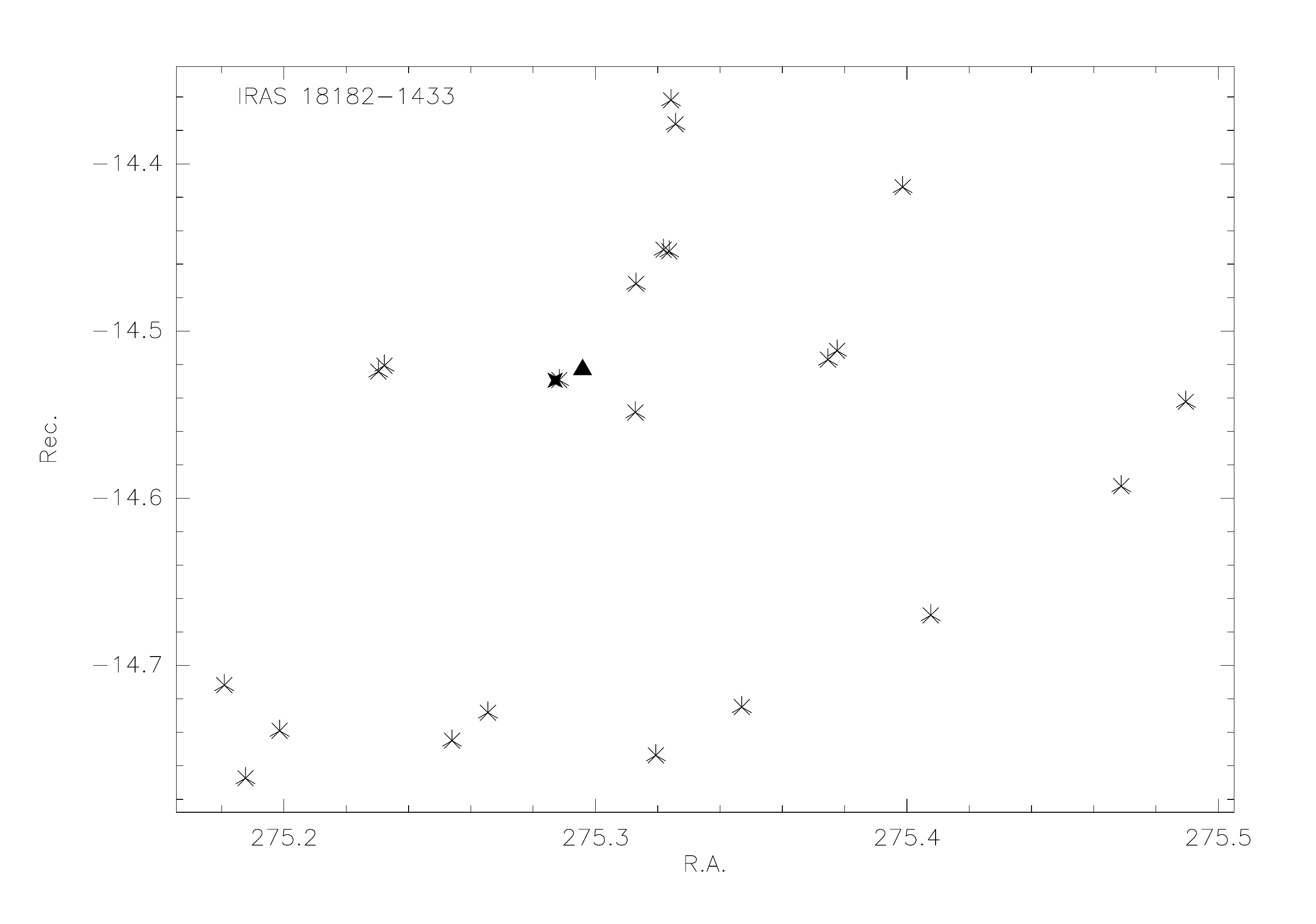}
  \includegraphics[angle=-90,width=7cm]{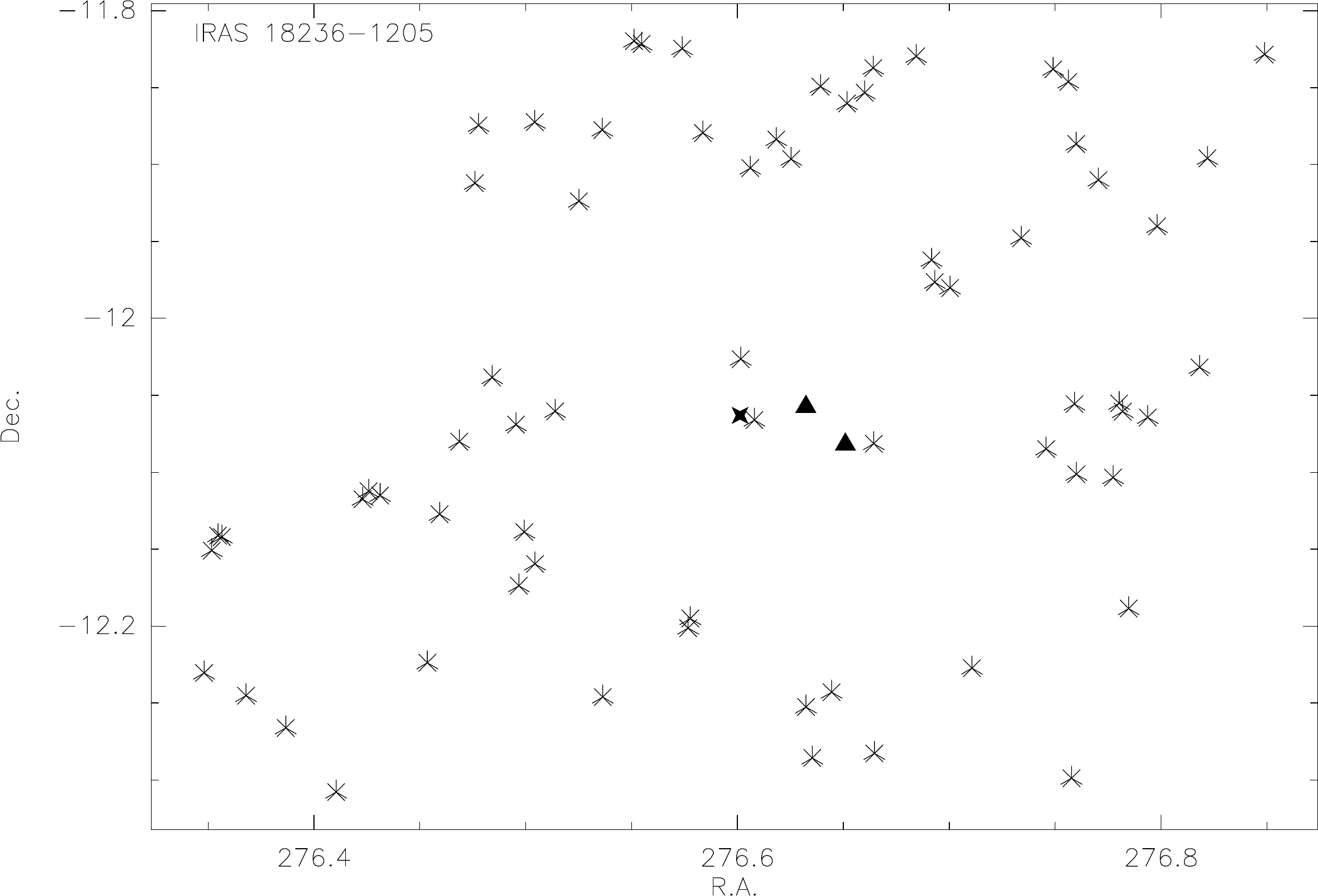}
}
%%========================================================================%%
  \resizebox{\hsize}{!}{
  \includegraphics[angle=-90,width=7cm]{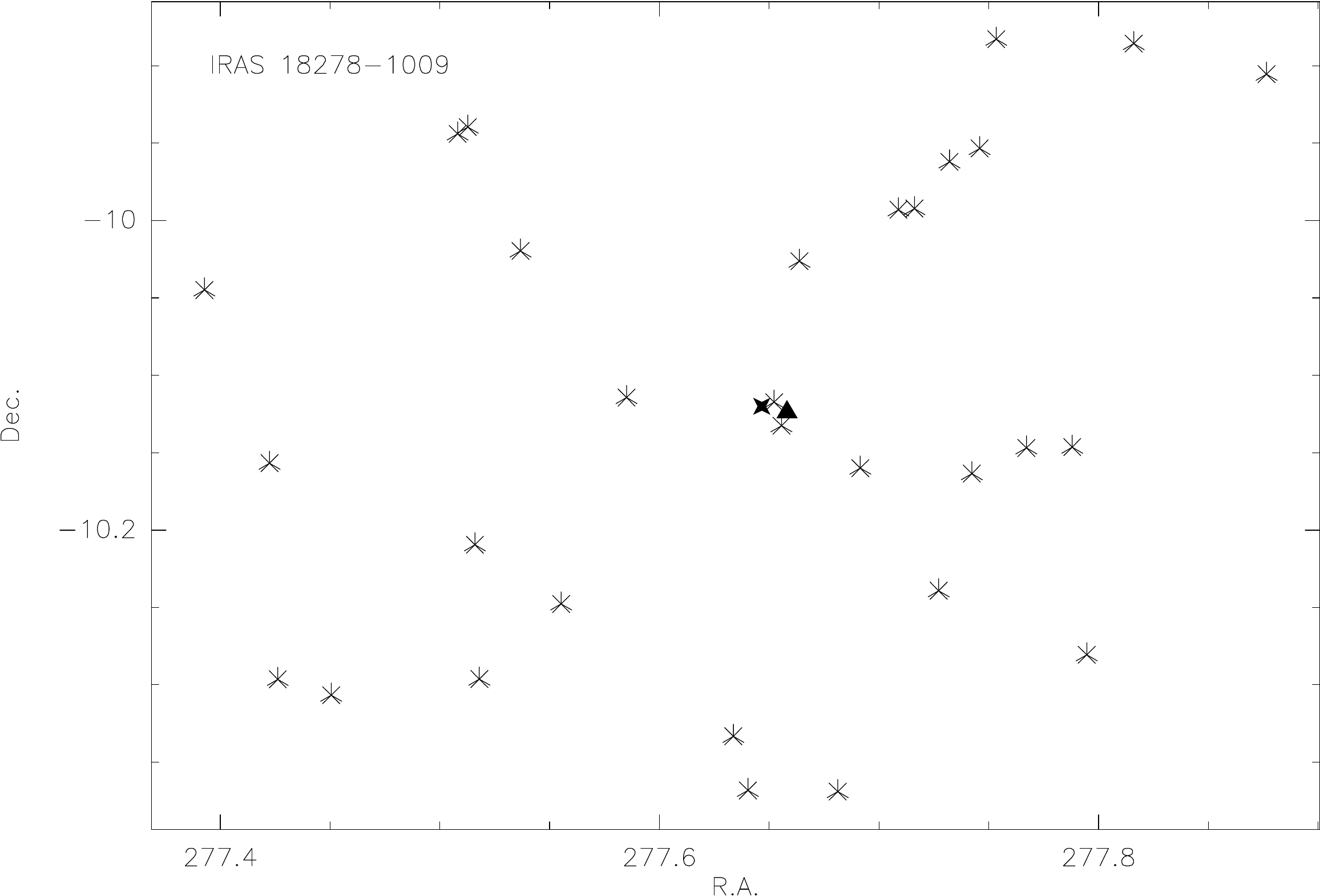}
  \includegraphics[angle=-90,width=7cm]{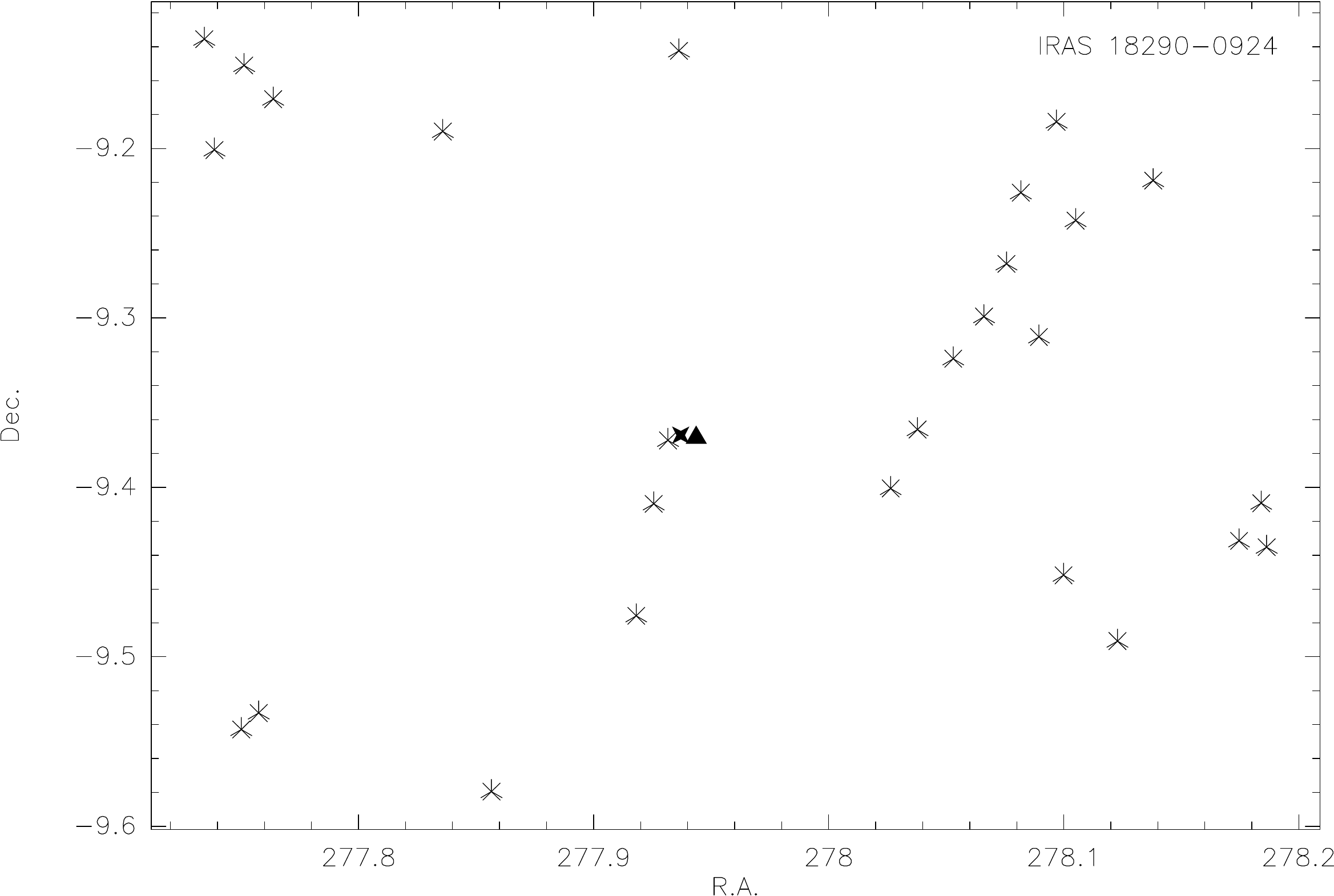}
}
%%========================================================================%%
  \resizebox{\hsize}{!}{
  \includegraphics[angle=-90,width=7cm]{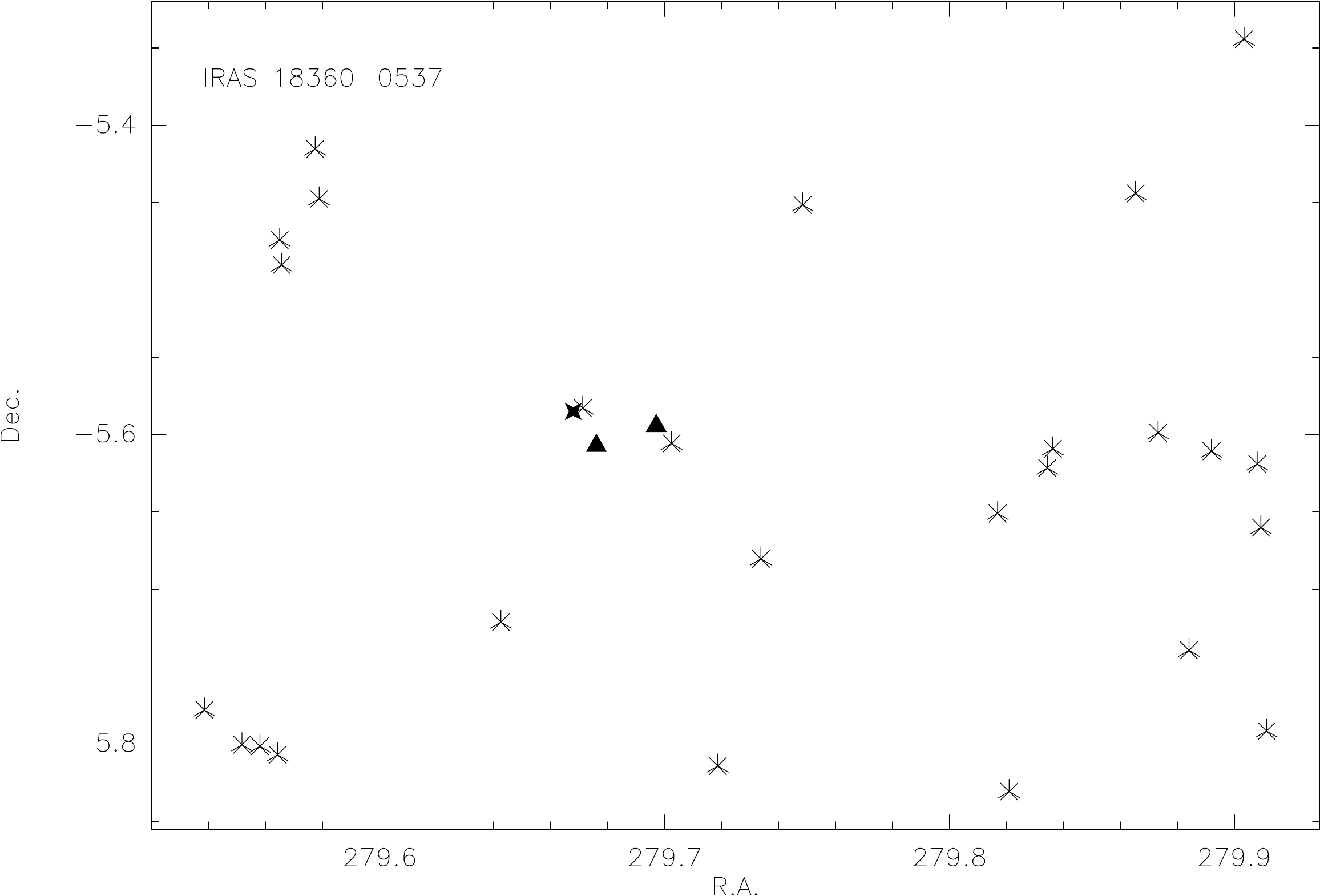}
  \includegraphics[angle=-90,width=7.5cm]{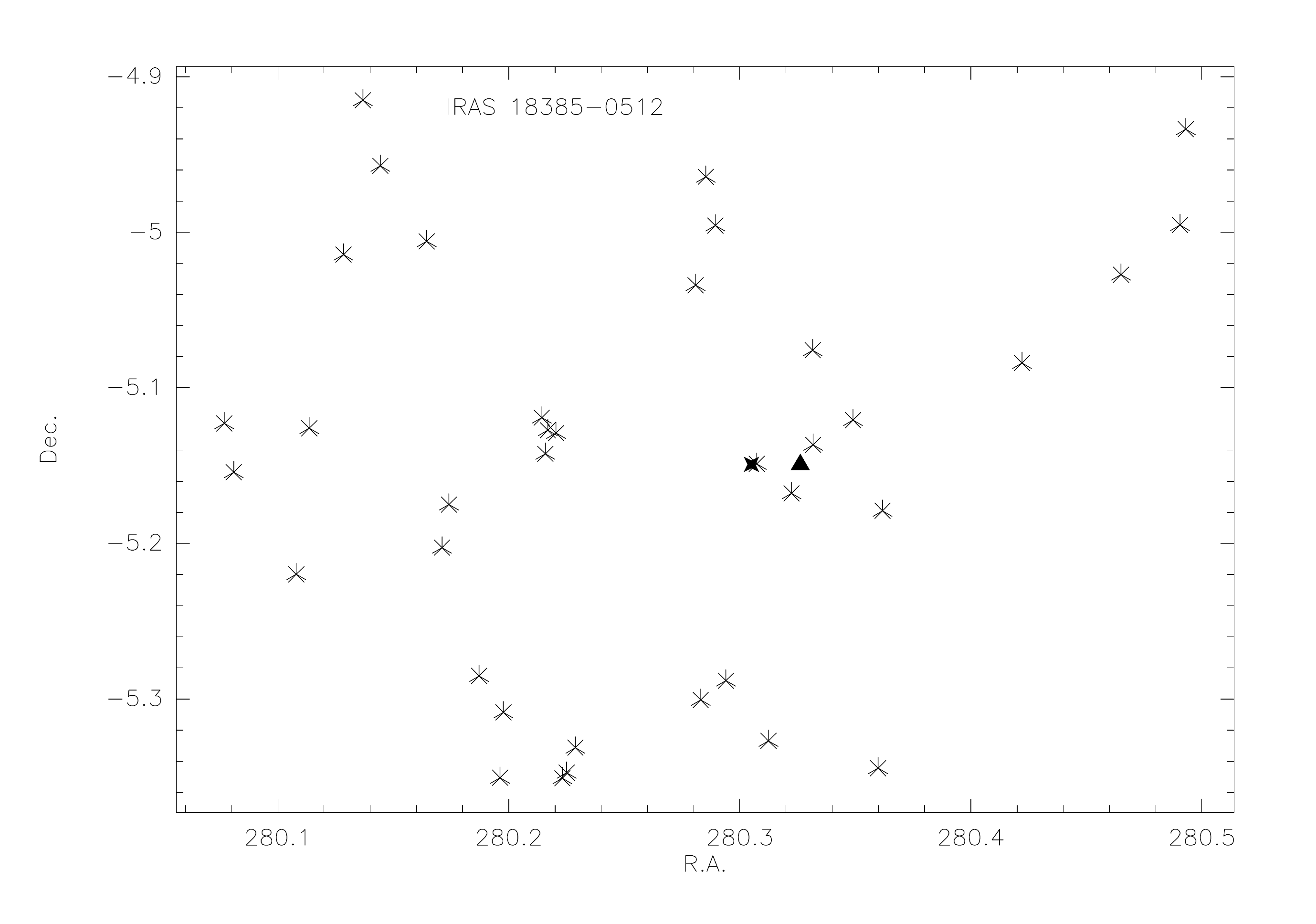}
}
%%========================================================================%%
  \resizebox{\hsize}{!}{
  \includegraphics[angle=-90,width=7.5cm]{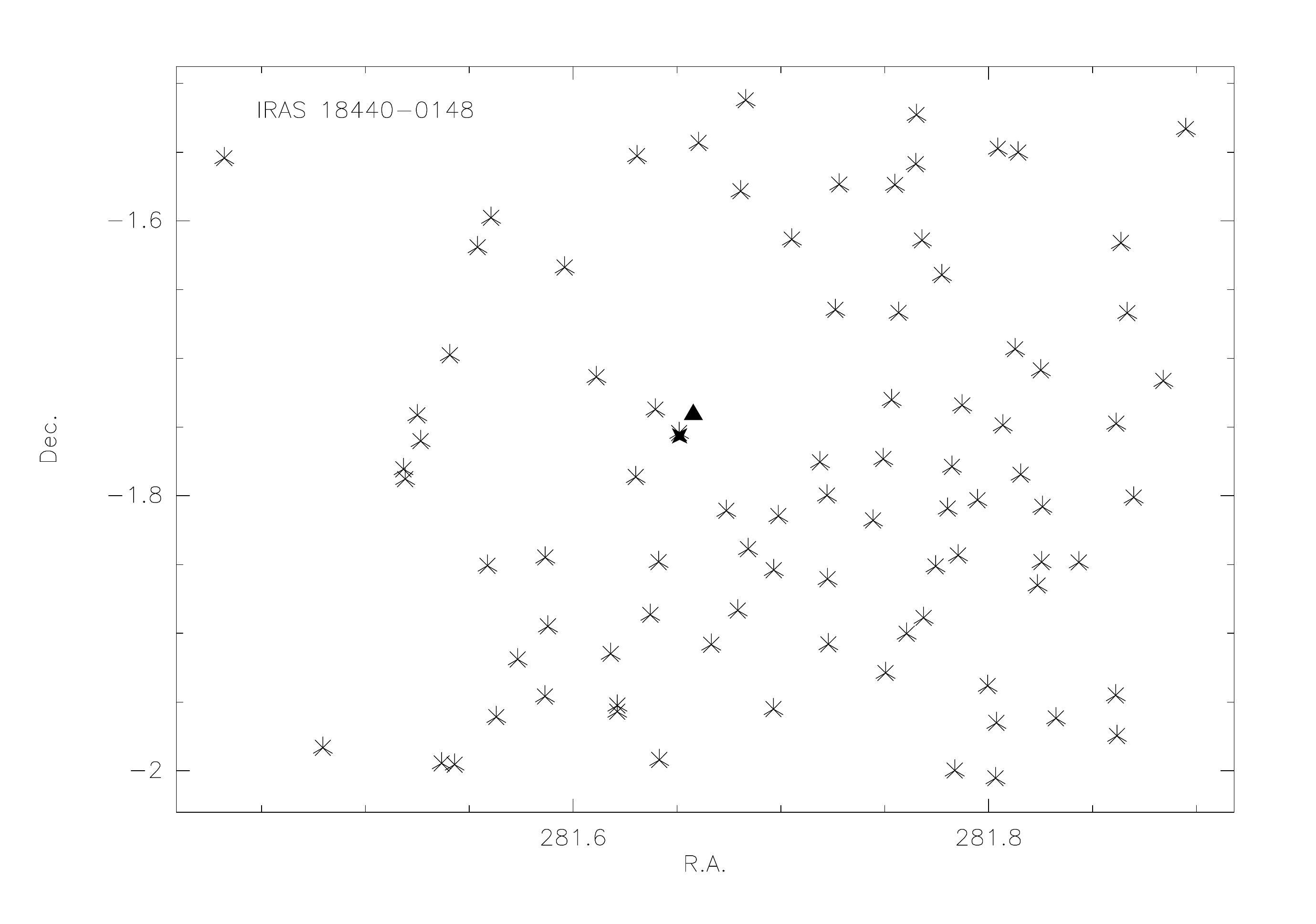}
  \includegraphics[angle=-90,width=7cm]{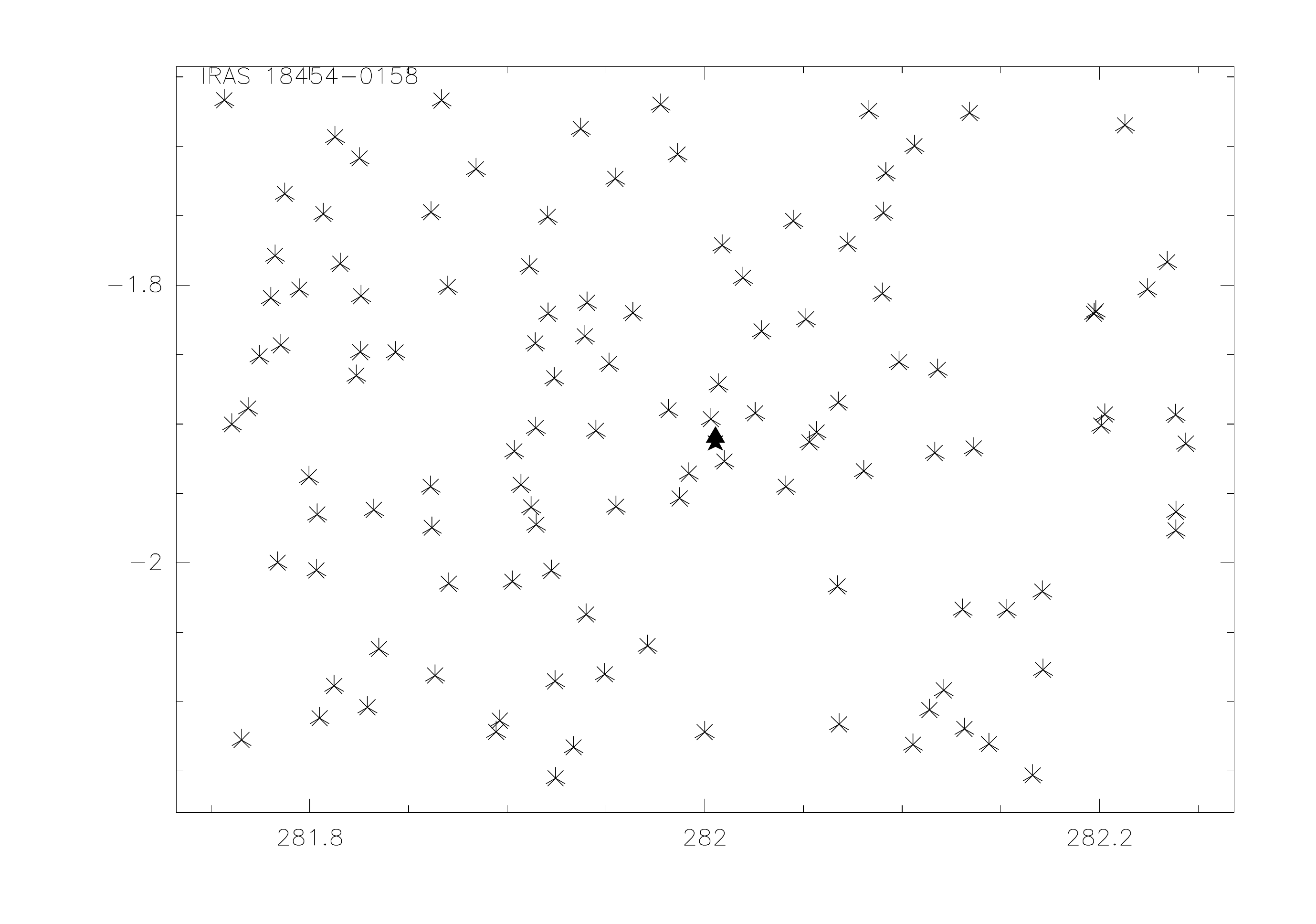}
}
%%========================================================================%%
\caption{same as Figure (\ref{fig:1mm}).}
%\label{fig:1mm}
  \end{figure*}

\end{document}